\documentclass[12pt]{article}
\usepackage{amsmath,amssymb,bm,graphicx}
\usepackage{color} 

\newcommand{\pslash}{\not \! p}

\newcommand{\Bslash}{\not \! B}
\newcommand{\tp}{\text{p}}
\newcommand{\tk}{\text{k}}
\newcommand{\sgn}{\text{sgn}}

\newcommand{\px}{{\bf p\cdot x }}
\numberwithin{equation}{section}
\usepackage{hyperref}
\usepackage{cite}

\begin{document}

%\vskip 0.5 truecm

\begin{center}
{\Large{\bf Quantum Field Theory of Particle Oscillations:\\\vskip 0.3cm Neutron-Antineutron Conversion}}
\end{center}\vskip .5 truecm
\begin{center}
{\bf \large{  Anca Tureanu}}

\vspace*{0.4cm} 
{\it {Department of Physics, University of Helsinki, P.O.Box 64, 
\\FIN-00014 Helsinki,
Finland
}}
\end{center}

\begin{abstract}
We formulate the quantum field theory description of neutron-antineutron oscillations in the framework of canonical quantization, in analogy with the Bardeen--Cooper--Schrieffer (BCS) theory and the Nambu--Jona-Lasinio model. The physical vacuum of the theory is a condensate of pairs of {\it would-be neutrons and antineutrons} in the absence of the baryon-number violating interaction.  The quantization procedure defines uniquely the mixing of massive Bogoliubov quasiparticle states which represent the neutron. In spite of not being mass eigenstates, neutron and antineutron states are defined on the physical vacuum and the oscillation formulated in asymptotic states. The exchange of baryonic number with the vacuum condensate engenders what may be observed as neutron-antineutron oscillation. 
The convergence between the present canonical approach and the Lagrangian/path integral approach to neutron oscillations is shown by the calculation of the anomalous (baryon-number violating) propagators. The quantization procedure proposed here can be extended to neutrino oscillations and, in general, to any particle oscillations.
\end{abstract}

\section{Introduction}\label{intro}

The Bardeen--Cooper--Schrieffer (BCS) theory of superconductivity\cite{BCS}, especially in Bogoliubov's treatment \cite{Bogoliubov}, became well known to particle physicists by the work of Nambu and Jona-Lasinio \cite{NJL}, who explored the analogy between the equations of motion governing the electrons in a superconductor near the Fermi level and the free Dirac equation of a massive fermion in Weyl representation. The spontaneous breaking of the $U(1)$ symmetry associated with  electric charge in the BCS theory is analogous to the spontaneous chiral symmetry breaking in the Nambu--Jona-Lasinio (NJL) model. Ever since, the BCS theory has influenced particle physics in a way that is hard to overestimate (see, e.g., Ref. \cite{Wilczek}).

Another analogy that can be established, this time in the language of Majorana fermions, is between the BCS Lagrangian and the effective Lagrangian of neutron-antineutron oscillations, which breaks baryon number symmetry \cite{kuzmin}-\cite{Rao-Shrock} (for a recent review, see, e.g., Ref.\cite{nnbar-review}). Neutron oscillations are a topical issue in present-day particle physics, mainly as a potentially observable window into the baryon-number violating phenomena that led to baryogenesis. Experimental searches  for neutron-antineutron conversion have been performed both with free neutron beams, and within nuclei \cite{exp}. At the European Spallation Source (ESS), new experiments  are being planned, aiming at improving by three orders of magnitude \cite{Milstead} the best bound on the oscillation time ($0.86\times10^8\ \text{s}$), obtained at ILL-Grenoble. Searches for neutron-mirror neutron oscillations \cite{Berezhiani-Bento} are also under consideration \cite{mirror_neutron_exp}. 
Recently, theoretical models have been proposed in which neutron-antineutron oscillations could occur moderately rapidly, at levels near to current limits and
within reach of an improved search, around $10^{9}-10^{10}$ s. Such models may be supersymmetric \cite{Babu-Mohapatra}, or involve large extra dimensions \cite{Nussinov-Shrock}, or still on constraints from post-sphaleron baryogenesis \cite{BDFM}.

Without speculating about the possible source of the baryon-number violation (which in principle can be achieved by spontaneous breaking of symmetry connected to a baryonic majoron \cite{Berezhiani-majoron}), the violation is usually considered to be explicit and not spontaneous. Nevertheless, Bogoliubov's formalism for the BCS theory can be adapted to the description of the neutron-antineutron oscillations, taking place via Bogoliubov quasiparticles of Majorana type, which represent the primary fermionic excitations of the system. The analogy between fermion oscillations with Majorana pseudoscalar mass term and BCS theory has been noted in the Lagrangian picture  in \cite{FT1, FT2}, where a relativistic equivalent of the Bogoliubov transformation (to which we shall return in Sect. \ref{Lagr_formalism}, eq. \eqref{RBT_def}) was used for the diagonalization of the Lagrangian. This intuitive connection is developed in this work into a general canonical quantization formulation.
The neutron and antineutron "states" are no more definite states in the physical Fock space of the system, however they can be defined as a superposition of (physical) mass eigenstates, by extension of the {\it would-be} neutron/antineutron states in the absence of the baryon-violating interaction. The collective excitations of quasiparticles, specific to BCS theory and NJL model, appear also in the baryon-number violating ground state of neutron oscillations. The condensate structure of vacuum for neutron and neutrino oscillations was first analyzed and explored in \cite{chang}. 

For compactness of terminology, we shall call {\it bare neutrons} the {\it would-be neutrons} in the absence of the baryon-number violating interaction. This is in analogy with the term {\it bare electron}, naming an electron without interaction with the lattice, in the BCS theory. In this language, the ground state of the baryon-number violating Hamiltonian is a condensate of pairs of bare neutrons and antineutrons, with opposite momenta and spins -- the analogues of Cooper pairs in the theory of superconductivity.

The approach described in this work can be applied as the quantum field theory of the free oscillations of any type of particles. Any oscillation phenomenon is a result of the fact that the fields that interact and appear in the relativistic construction of the Lagrangian are not fields with definite mass, but mixings thereof. On the other hand, the massive states are not observed individually, but only in time-dependent superpositions, representing the oscillating "particles". The mixing is caused by some additional interactions (baryon-number violating interaction in the case of neutrons, lepton number violating in the case of neutrinos, strangeness violating weak interaction in the case of $K_0$ mesons etc.), which are not taken into account when the oscillating particles are produced. The vacuum condensate is then the reservoir of fermionic number, strangeness, etc., as well as of the extra chirality, which gives their definite masses but indefinite quantum numbers to the quasiparticles associated with  the mass eigenfields. The difference in the masses of the quasiparticles produces the oscillation, which is essentially the oscillation of a quantum number, realized through the exchange with the vacuum condensate. We may say that the quasiparticles as mass eigenstates create the observable kinematical effects, while the oscillating particles are subject to the dynamical effects included in the Lagrangian. The interaction prevents the mass discrimination and the lack of a "mass analyser" leads to what is perceived as particle oscillations.

The neutrino oscillations are the prototypical oscillation phenomena that have been studied extensively over a long time (for a review and a historical account, see Ref. \cite{Bilenky_hist}). The quantum field theoretical framework has been developed and the prevailing picture consists in viewing the neutrino oscillations as a single process encompasing production, propagation and detection, with the neutrino in the intermediate (virtual) state. This approach was pioneered in Ref. \cite{GKLL} (see also the reviews \cite{neutrino_rev} and references therein for an updated status). In spite of the concentrated effort and several ingenious theoretical solutions, there are paradoxical features \cite{Akhmedov_Smirnov} and questions on whose answer there is still no consensus, such as:  how are the flavour neutrino states supposed to be defined? are the flavour states momentum eigenstates? is it necessary that the massive neutrinos which mix have equal energies?  are the flavour states physical and in which sense? These are fundamental issues arising about any oscillation process.

The quantization method described in this work provides a natural  and unequivocal answer to a crucial question: if we know that a certain field is expressed as a definite mixing of other fields, how do we define the states corresponding to the former field in terms of the states of the latter fields? The answer will be given in Sect. \ref{trans_prob}, and it will turn out to have been impossible to guess without invoking the power of the present quantization procedure.

\section{Lagrangian description of neutron-antineutron oscillations}\label{Lagr_formalism}

The free neutron-antineutron oscillations are analyzed by the quadratic effective Hermitian Lagrangian with general $\Delta B=2$ terms added:
\begin{eqnarray}\label{1}
{\cal L}&=&\overline{\Psi}(x)i\gamma^{\mu}\partial_{\mu}\Psi(x) - m\overline{\Psi}(x)\Psi(x)\nonumber\\
&-&\frac{1}{2}\epsilon_{1}[\Psi^{T}(x)C\Psi(x) +\overline{\Psi}(x)C\overline{\Psi}^{T}(x)]\nonumber\\
&-&\frac{i}{2}\epsilon_{5}[e^{-i\alpha}\Psi^{T}(x)C\gamma_{5}\Psi(x)  +e^{i\alpha} \overline{\Psi}(x)C\gamma_{5}\overline{\Psi}^{T}(x)],
\end{eqnarray}
where $m$, $\epsilon_{1}$, $\epsilon_{5}$ and $\alpha$ are real parameters, $\Psi(x)$ is the neutron field and $C$ is the charge-conjugation matrix. The Standard Model Lagrangian is invariant under the global $U(1)$ baryonic number transformations, under which the neutron field transforms as
$\Psi(x)\to e^{i\beta}\Psi(x).$
Clearly, under such a transformation the terms proportional to $\epsilon_1$ and $\epsilon_5$ in the Lagrangian \eqref{1} are noninvariant.
They are the only Lorentz-invariant baryon-number violating terms that can be written and they are Majorana mass terms of scalar and pseudoscalar type. A pseudoscalar mass term $im'\overline{\Psi}(x)\gamma_5\Psi(x)$ can in principle be added as well, but we shall omit it, as its role in this $\Delta B=2$ Lagrangian is supplanted by the $\epsilon_5$ term. The baryon-number violating terms are quadratic, but it is assumed that they are the effective expression of an interaction whose details are unknown. For this reason, when referring to those terms we may use the  term baryon-number violating interaction.

We shall adopt a charge conjugation invariant version of the Lagrangian \eqref{1}. With the traditional convention for defining the charge conjugated spinor as 
\begin{eqnarray}\label{C-def}
{\cal C}\Psi(x){\cal C}^{-1}=\Psi^{c}(x)=C\bar{\Psi}^{T}(x),
\end{eqnarray}
the Lagrangian \eqref{1} is invariant under charge conjugation only when $\alpha=0$, therefore we fix the phase in this way. Irrespective of the phase $\alpha$, the Lagrangian \eqref{1} is parity violating. This can be easily seen if we adopt the convention
\begin{eqnarray}\label{P-def}
{\cal P}\Psi({\bf x},t){\cal P}^{-1}=\gamma_0{\Psi}(-{\bf x},t).
\end{eqnarray}
There is no phase convention for the definition of parity that can render the Lagrangian invariant. Thus, the Lagrangian
\begin{eqnarray}\label{Lagr_osc}
{\cal L}&=&\overline{\Psi}(x)i\gamma^{\mu}\partial_{\mu}\Psi(x) - m\overline{\Psi}(x)\Psi(x)\nonumber\\
&-&\frac{1}{2}\epsilon_{1}[\Psi^{T}(x)C\Psi(x) +\overline{\Psi}(x)C\overline{\Psi}^{T}(x)]\nonumber\\
&-&\frac{i}{2}\epsilon_{5}[\Psi^{T}(x)C\gamma_{5}\Psi(x)  +\overline{\Psi}(x)C\gamma_{5}\overline{\Psi}^{T}(x)]
\end{eqnarray}
is C-invariant and P- and CP-violating. The P- and CP-violation are in line with the expected electric dipole moment for the neutron and cannot be eliminated from the Lagrangian by any field redefinition\footnote{It was shown in \cite{FT1} that the partial diagonalization of the Lagrangian \eqref{Lagr_osc}, which removes the baryon-number violating $\epsilon_5$ term, leads to a term of the type $im'\overline{\Psi}(x)\gamma_5\Psi(x)$, which may reflect the effect of the QCD $\theta$-vacuum.}. The effect is due to the interplay of the "vector coupling" in the $\epsilon_1$ term and the "axial vector coupling" in the $\epsilon_5$ term. Incidentally, if in the Lagrangian \eqref{1} one takes either $\epsilon_1=0$ or $\epsilon_5=0$, the remaining Lagrangian can always be shown to be both P- and C-invariant, by a redefinition of the phase of the corresponding operations. 

The diagonalization of the Lagrangian \eqref{1} and the analysis of neutron-antineutron oscillations were performed in detail in Ref. \cite{FT1}. The P- and CP-violation of the Lagrangian were shown not to have observable effects in the free $n-\bar n$ transition probability (for recent discussions of the discrete symmetries, especially CP, in neutron-antineutron oscillations, see \cite{FT1}-\cite{gardner}). The Lagrangian analysis in \cite{FT1} involved the introduction of a relativistic analogue of the Bogoliubov transformation, which mixes the fields $\Psi(x)$ and $\Psi^c(x)$, and diagonalizes the $\epsilon_5$ term:
\begin{eqnarray}\label{RBT_def}
\left(\begin{array}{c}
            \Psi(x)\\
            \Psi^{c}(x)
            \end{array}\right)
&=& \left(\begin{array}{c}
            \cos\Theta N(x)-i\gamma_{5}\sin\Theta N^c(x)\\
            \cos\Theta N^c(x)-i\gamma_{5}\sin\Theta N(x)
            \end{array}\right),
\end{eqnarray}
where the fields $N, N^c$ are of Dirac type and $\sin2\Theta=\epsilon_4/\sqrt{m^2+\epsilon_5^2}$.
 The role of this transformation is crucial in the exact Lagrangian analysis of neutron oscillation. Also, in the context of the seesaw mechanism, which is described by the Lagrangian \eqref{1} with $\alpha=\pi/2$, a similar (but C-noninvariant) relativistic Bogoliubov transformation proves essential in absorbing the C-violation and rendering the Majorana neutrino a proper eigenfield of the charge conjugation operator on a new vacuum \cite{FT,FT2}. We shall return to the details of the relativistic Bogoliubov transformation in Sect. \ref{propagator}, when comparing the results of the Lagrangian and Hamiltonian formulations in the calculation of the anomalous (baryon-number violating) propagators.

Here, we collect only a few necessary formulas pertaining to the Lagrangian formalism, which will be needed later on. The equations of motion derived from the Lagrangian \eqref{Lagr_osc} are:
\begin{eqnarray}\label{eom_Psi}
&&(i\gamma^{\mu}\partial_{\mu}-m)\Psi(x)-(\epsilon_1+i\epsilon_5\gamma_5) \Psi^{c}(x)=0,\nonumber\\
&&(i\gamma^{\mu}\partial_{\mu}-m)\Psi^{c}(x)-(\epsilon_1+i\epsilon_5\gamma_5) \Psi(x)=0,
\end{eqnarray}
with $\Psi^{c}=C\overline{\Psi}^{T}$. We rewrite them as 
\begin{eqnarray}\label{EOM_gen}
(i\gamma^{\mu}\partial_{\mu}-(m+\epsilon_1)-i\epsilon_5\gamma_5)\Psi_+(x)=0,\cr
(i\gamma^{\mu}\partial_{\mu}-(m-\epsilon_1)+i\epsilon_5\gamma_5) \Psi_-(x)=0,
\end{eqnarray}
in terms of the Majorana fields
\begin{eqnarray}\label{Majorana}
\Psi_{\pm}(x)=\frac{1}{\sqrt2}\big(\Psi(x)\pm \Psi^{c}(x)\big),
\end{eqnarray}
which satisfy Dirac equations with {\em different masses} and thus diagonalize the Lagrangian \eqref{Lagr_osc}. The mass eigenvalues are easily obtained by setting $\Psi_\pm(x)=e^{ipx}\Psi_\pm(p)$:
\begin{eqnarray}\label{EOM_p}
(\pslash-(m+\epsilon_1)-i\epsilon_5\gamma_5)\Psi_+(p)=0,\cr
(\pslash-(m-\epsilon_1)+i\epsilon_5\gamma_5) \Psi_-(p)=0.
\end{eqnarray}
For $\Psi_+(p)$ we note that
$$\pslash-(m+\epsilon_1)-i\epsilon_5\gamma_5=0,$$
which is rewritten as
$$\pslash-i\epsilon_5\gamma_5=m+\epsilon_1.$$
From here we find, for $\Psi_+(p)$,
\begin{eqnarray}\label{M+}
p^2=M_+^2=(m+\epsilon_1)^2+\epsilon_5^2,
\end{eqnarray}
while for $\Psi_-(p)$ we obtain
\begin{eqnarray}\label{M+}
p^2=M_-^2=(m-\epsilon_1)^2+\epsilon_5^2.
\end{eqnarray}
Upon diagonalization, the Lagrangian \eqref{Lagr_osc} becomes
\begin{eqnarray}\label{Lagr_diag}
{\cal L}&=&\frac{1}{2}\left[\overline{\Psi}_+(x)i\gamma^{\mu}\partial_{\mu}\Psi_+(x) - M_+\overline{\Psi}_+(x)\Psi_+(x)\right]\cr
&+&\frac{1}{2}\left[\overline{\Psi}_-(x)i\gamma^{\mu}\partial_{\mu}\Psi_-(x) - M_-\overline{\Psi}_-(x)\Psi_-(x)\right].
\end{eqnarray}

Thus, the Lagrangian description leads to the expression of the neutron field $\Psi(x)$ as mixing of the free massive Majorana fields $\Psi_\pm(x)$:
\begin{eqnarray}\label{neutron_field_mixing}
\Psi(x)=\frac{1}{\sqrt2}\big(\Psi_+(x)+ \Psi_-(x)\big).
\end{eqnarray}
The rest of this paper will be concerned with finding the superposition of states of $\Psi_\pm$ which represent the states corresponding to the field $\Psi$. For this purpose, we shall have to pass to the Hamiltonian description of the model.

In anticipation, let us still recall that a proper Dirac field $\psi(x)$ of mass $m$ can always be written as the sum of two Majorana fields with the same mass $m$, which are constructed as $\psi_{\pm}(x)=\frac{1}{\sqrt2}(\psi(x)\pm \psi^{c}(x))$. The neutron field $\Psi(x)$, however, is the mixture of two mass-nondegenerate Majorana fields, therefore not a Dirac field. The meaning of the neutron and antineutron as "particle states" associated with the field $\Psi(x)$ becomes more subtle.

\section{Hamiltonian description and vacuum structure}\label{Hamilt_formalism}

The direct way to canonically quantize the model described by the Lagrangian \eqref{Lagr_osc} is by solving the equations of motion \eqref{EOM_p}, for the Majorana fields with definite masses, and applying equal-time canonical anticommutators, which would lead to the algebra of the creation and annihilation operators. The system is exactly solvable. However, such a straightforward method would obscure the baryon-number violation, as well as the associated dynamical mass generation. For this reason, we shall adopt a different method of canonical quantization, which has the benefit of uncovering a more telling intuitive picture of the oscillation phenomenon.

\subsection{Canonical quantization and Bogoliubov quasiparticles}\label{canon_quant}

The method which we are going to use is analogous to the one developed by Bogoliubov for the treatment of the BCS model \cite{Bogoliubov} (for a pedagogical presentation, see, e.g., Ref. \cite{monograph_cond_mat}) and by Nambu and Jona-Lasinio in their BCS-inspired theory of dynamical generation of nucleon masses \cite{NJL} (see also \cite{UTK}). In current parlance, it is based on the unitarily inequivalent representations of canonical (anti)commutators.  An ample exposition thereof  can be found in the monograph \cite{Umezawa-book}. Unitarily inequivalent representations can exist only in systems with infinite number of degrees of freedom, in other words in quantum field theory. In constrast, in quantum mechanics, the Stone--von Neumann theorem ensures that all representations of the canonical commutators are unitarily equivalent. The existence of unitarily inequivalent representations in quantum field theory is an essential ingredient of Haag's theorem \cite{Haag}. Here, we shall summarize the main aspects needed for the application to the neutron-antineutron oscillation model.

In the Heisenberg picture, the evolution of a fermionic system is expressed by the time-dependent Heisenberg fields, satisfying the equation of motion
\begin{equation}
i\partial_t\Psi(x)=[\Psi(x),H],
\end{equation}
where $H$ is the Hamiltonian of the system and the fields $\Psi(x)$ satisfy equal-time anticommutation relations. The Heisenberg fields act on the Fock space of physical states, i.e. those states that correspond to observable free particles. They are obtained by  the application of creation operators to the physical vacuum of the model. Consequently, the Fock space has to satisfy the requirement that the Heisenberg fields are expressed in terms of creation and annihilation operators of the physical free particles. (It is perhaps more familiar to think that the Heisenberg fields are expressed by a Dyson expansion in terms of incoming or outgoing physical fields). When this condition is fulfilled, the total Hamiltonian of the system takes the form of a free Hamiltonian. This is one of the essential features of the Heisenberg picture and will provide us the basis for solving the baryon-number violating model defined by the Lagrangian \eqref{Lagr_osc}. The method described below is called the {\it self-consistent method}, in the sense that it relies on the self-consistency between the Hamiltonian and the choice of the Fock space of physical particles \cite{Umezawa-book}.

In practice, we obtain first the classical Hamiltonian of the system starting from the Lagrangian. Then we choose a candidate for the  physical field (i.e. a field which satisfies a certain free field equation of motion) and quantize it canonically. We go to the Schr\"odinger picture and express the Hamiltonian in terms of the creation and annihilation operators of the candidate field. It is clear that there are infinitely many free fields to choose from, each defined by a different mass. Typically, one makes a meaningful selection by considering the solution of the free part $H_0$ of the total Hamiltonian $H$ (see, for example, Ref. \cite{Bog-Shirk}). If the Hamiltonian is not diagonal, then the candidate field is not the physical field. What we need to do is to diagonalize the Hamiltonian. Upon diagonalization, the Hamiltonian will be expressed in terms of the creation and annihilation operators of the true physical fields, acting on the true ground state of the model. The physical Fock space can be constructed and the problem is solved (the return to the Heisenberg picture being then straightforward). The diagonalization is achieved by establishing certain relations between the creation and annihilation operators of the candidate field and those of the true physical field. These turn out to be Bogoliubov transformations, preserving the canonicity of the algebra. The quanta of the physical fields will therefore be Bogoliubov quasiparticles. The two sets of canonical operators are related by a transformation which seemingly is unitary. However, it turns out that they act {\it as creation and annihilation operators} in two orthogonal Fock spaces, constructed on orthogonal vacua, therefore the transformation is not unitarily implementable \cite{Umezawa-book}. For this reason, the two Fock spaces are said to be unitarily inequivalent representations of the canonical algebra. 

Although there are in principle an infinity of unitarily inequivalent representations, it should be stressed once more that only one representation is physical, and that is the one in which the Hamiltonian is diagonal. The corresponding vacuum is the one and only vacuum of the theory (except the case when there is spontaneous breaking of symmetry).

All the assertions in the above summary will be substantiated below on the concrete model defined by the Lagrangian \eqref{Lagr_osc}. We shall also draw some parallels with the BCS theory or the NJL model whenever these parallels may prove illuminating.

\subsubsection*{Hamiltonian}

We start by writing the Hamiltonian corresponding to the baryon-number violating Lagrangian \eqref{Lagr_osc}:
\begin{eqnarray}\label{Hamilt_osc}
{H}&=&\int d^3x\left(-
\overline{\Psi}(x)i\gamma^{k}\partial_{k}\Psi(x) + m\overline{\Psi}(x)\Psi(x)\right)\cr
&+&\int d^3x\ \frac{\epsilon_{1}}{2}\left(\Psi^{T}(x)C\Psi(x) + \overline{\Psi}(x)C\overline{\Psi}^{T}(x)\right)\cr
&+&\int d^3x\ i\frac{\epsilon_{5}}{2}\left(\Psi^{T}(x)C\gamma_{5}\Psi(x) x +\overline{\Psi}(x)C\gamma_{5}\overline{\Psi}^{T}(x)\right),\cr
&=&H_0 +H_{\Bslash},
\end{eqnarray}
where $H_0$ stands for the Dirac Hamiltonian of a field with thes mass $m$, while $H_{\Bslash}$ represents the baryon-number violating part.

\subsubsection*{Choice of a candidate physical field}

The next step is to pick up a candidate $\psi(x)$ for the r\^ole of physical field. The meaningful choice out of the arbitrary possibilities is to take $\psi(x)$ as the {\it would-be neutron field} in the absence of the baryon-number violating interaction. This is
the solution of the free Dirac equation with mass $m$,
\begin{eqnarray}\label{Dirac_eq}
&&(i\gamma^{\mu}\partial_{\mu}-m)\psi(x)=0,
\end{eqnarray}
in other words, the eigenfield of the free Dirac Hamiltonian $H_0$ in \eqref{Hamilt_osc}.
Hence, we proceed by going to the Schr\"odinger picture, at $t=0$, and making the identification \cite{Bogoliubov, UTK,Bog-Shirk}
\begin{equation}\label{NJL}
\Psi({\bf x},0)=\psi({\bf x},0),
\end{equation}
in the Hamiltonian \eqref{Hamilt_osc}.
In this way, we can naturally assign baryonic quantum numbers to the quanta of the field $\psi(x)$, which will be called {\it bare neutrons and antineutrons}. Moreover, in the limit $\epsilon_1,\epsilon_5\to0$, the states associated with the field $\Psi(x)$ coincide with the states associated with $\psi(x)$. Consequently, we shall have a handle to define what is meant by neutron and antineutron when baryonic number is violated.

We expand the Hamiltonian  \eqref{Hamilt_osc} in terms of the modes of the bare neutron field,
\begin{eqnarray}\label{Dirac_mode_exp}
\psi({\bf x},0)=\int\frac{d^3p}{(2\pi)^{3/2}\sqrt{2\omega_\tp}}\sum_\lambda\left(a_\lambda({\bf p})u_\lambda({\bf p})e^{i{\bf p\cdot x }}+b^\dagger_\lambda({\bf p})v_\lambda({\bf p})e^{-i{\bf p\cdot x }}\right),
\end{eqnarray}
which is written in helicity basis (see Appendix \ref{appendix1}), with $\omega_\tp=\sqrt{\tp^2+m^2}$.
The charge conjugated spinor $\psi^c=C\bar\psi^T$, with the conventions from Appendix \ref{appendix1}, is
\begin{eqnarray}\label{charge_conj}
\psi^c({\bf x},0)=\int\frac{d^3p}{(2\pi)^{3/2}\sqrt{2\omega_\tp}}\sum_\lambda\text{sgn}\,\lambda\left(b_\lambda({\bf p})u_\lambda({\bf p})e^{i{\bf p\cdot x }}+a^\dagger_\lambda({\bf p})v_\lambda({\bf p})e^{-i{\bf p\cdot x }}\right).
\end{eqnarray}
The operators $a, a^\dagger, b, b^\dagger$ are creation and annihilation operators on a vacuum $|0\rangle$, which we may call {\it particle vacuum},
\begin{eqnarray}\label{naive_vac}
a_\lambda({\bf p})|0\rangle=b_\lambda({\bf p})|0\rangle=0\end{eqnarray}
and satisfy ordinary anticommutation relations:
\begin{eqnarray}\label{ACR_ord}
\{a_\lambda({\bf p}),a^\dagger_{\lambda'}({\bf k})\}&=&\delta_{\lambda\lambda'}\delta({\bf p}-{\bf k}),\\
\{b_\lambda({\bf p}),b^\dagger_{\lambda'}({\bf k})\}&=&\delta_{\lambda\lambda'}\delta({\bf p}-{\bf k}),\nonumber
\end{eqnarray}
all the other anticommutators being zero. The states 
\begin{eqnarray}\label{bare_neutron}
a^\dagger_\lambda({\bf p})|0\rangle \ \ \ \text{and }\ \ \ \ b^\dagger_\lambda({\bf p})|0\rangle\end{eqnarray}
represent bare neutron and antineutron states, respectively, of mass $m$ and definite momentum and helicity. We assign baryonic number $+1$ to the bare neutron states and $-1$ to the bare antineutron states. In analogy with the theory of neutrino oscillations, we may think about the Fock space of states built on the vacuum $|0\rangle$ as a space of {\it flavour states}. 

\subsubsection*{Mode expansion of the Hamiltonian}

Using \eqref{Dirac_mode_exp} and \eqref{charge_conj} in \eqref{Hamilt_osc}, we find, with the help of the relations \eqref{spinor_rel}:
\begin{eqnarray}\label{H_Dirac_modes}
H_0=\int d^3 p \sum_{\lambda}\omega_\tp\left(a^\dagger_\lambda({\bf p}) a_\lambda({\bf p})+b^\dagger_\lambda({\bf p})b_\lambda({\bf p})\right)
\end{eqnarray}
and
\begin{eqnarray}\label{H_b_viol}
H_{\Bslash}&=&\int d^3 p \sum_{\lambda}\Big[\epsilon_1\frac{m}{\omega_\tp}\text{sgn}\,\lambda\, \left( a^\dagger_\lambda({\bf p}) b_\lambda({\bf p})+b^\dagger_\lambda({\bf p})a_\lambda({\bf p})\right)\\
&-&i\epsilon_1\frac{\text{p}}{2\omega_\tp}\left(a_\lambda({\bf p}) a_\lambda(-{\bf p})+a^\dagger_\lambda({\bf p}) a^\dagger_\lambda(-{\bf p})+b_\lambda({\bf p}) b_\lambda(-{\bf p})+b^\dagger_\lambda({\bf p})b^\dagger_\lambda(-{\bf p})\right)\cr
&+&\sgn\,\lambda\frac{\epsilon_5}{2}\left(a_\lambda({\bf p}) a_\lambda(-{\bf p})-a^\dagger_\lambda({\bf p}) a^\dagger_\lambda(-{\bf p})+b_\lambda({\bf p}) b_\lambda(-{\bf p})-b^\dagger_\lambda({\bf p})b^\dagger_\lambda(-{\bf p})\right)\Big].
\nonumber
\end{eqnarray}
The baryon-number violating part of the Hamiltonian is, as expected, non-diagonal. The terms containing $a^\dagger_\lambda({\bf p}) b_\lambda({\bf p})+b^\dagger_\lambda({\bf p})a_\lambda({\bf p})$ indicate the neutron-antineutron transition. The rest of the non-diagonal terms suggest the pairing of neutrons and antineutrons, in the manner of the Cooper pairs in the BCS theory \footnote{In the BCS theory, the Hamiltonian is written in terms of the creation and annihilation operators of the bare electrons, the interaction with the lattice providing the nondiagonal terms.}. We omit the vacuum energy and present throughout the Hamiltonian in normal form.

We may proceed from here to the diagonalization, but it is technically advantageous to take into account the hint provided by the equations of motion \eqref{EOM_gen}, namely that the fields which diagonalize the Lagrangian are Majorana fields. Therefore, we shall re-express the Hamiltonian \eqref{Hamilt_osc} in terms of the creation and annihilation operators associated with the degenerate Majorana fields of mass $m$ into which the Dirac field $\psi(x)$ can be split.

We note that the convention adopted for the charge conjugation transformation leads to the following action on the creation and annihilation operators:
\begin{eqnarray}
{\cal C}a_\lambda({\bf p}){\cal C}^{-1}= \text{sgn}\,\lambda\, b_\lambda({\bf p}),\ \ \ \ {\cal C}b_\lambda({\bf p}){\cal C}^{-1}= \text{sgn}\,\lambda \, a_\lambda({\bf p}).
\end{eqnarray}
As a result, we obtain the creation and annihilation operators of the Majorana fields $\psi_\pm(x)$ defined by
\begin{eqnarray}\label{Majorana_cand}
\psi_{\pm}(x)=\frac{1}{\sqrt2}\big(\psi(x)\pm \psi^{c}(x)\big),
\end{eqnarray}
identified by the subscript $_M$, in the form:
\begin{eqnarray}\label{M-D}
a_{M\lambda}({\bf p})=\frac{1}{\sqrt2}\left(a_\lambda({\bf p})+ \text{sgn}\,\lambda\, b_\lambda({\bf p})\right),\cr
b_{M\lambda}({\bf p})=\frac{1}{\sqrt2}\left(a_\lambda({\bf p})- \text{sgn}\,\lambda\, b_\lambda({\bf p})\right).
\end{eqnarray}
The inverse of the above transformation reads:
\begin{eqnarray}\label{D-M}
a_{\lambda}({\bf p})=\frac{1}{\sqrt2}\left(a_{M\lambda}({\bf p})+ b_{M\lambda}({\bf p})\right),\cr
\text{sgn}\,\lambda\, b_{\lambda}({\bf p})=\frac{1}{\sqrt2}\left(a_{M\lambda}({\bf p})-  b_{M\lambda}({\bf p})\right).
\end{eqnarray}
At $t=0$, the Majorana fields $\psi_\pm({\bf x},0)$ are expressed as:
\begin{eqnarray}\label{Majorana_mode_exp}
&\psi_+({\bf x},0)=\int\frac{d^3p}{(2\pi)^{3/2}\sqrt{2\omega_\tp}}\sum_\lambda\left(a_{M\lambda}({\bf p})u_\lambda({\bf p})e^{i{\bf p\cdot x }}+\sgn\,\lambda\ a^\dagger_{M\lambda}({\bf p})v_\lambda({\bf p})e^{-i{\bf p\cdot x }}\right),\cr
&\psi_-({\bf x},0)=\int\frac{d^3p}{(2\pi)^{3/2}\sqrt{2\omega_\tp}}\sum_\lambda\left(b_{M\lambda}({\bf p})u_\lambda({\bf p})e^{i{\bf p\cdot x }}-\sgn\,\lambda\ b^\dagger_{M\lambda}({\bf p})v_\lambda({\bf p})e^{-i{\bf p\cdot x }}\right).
\end{eqnarray}

Using the formulas \eqref{D-M} we recast the Hamiltonian \eqref{Hamilt_osc} in terms of the Majorana operators:
\begin{eqnarray}\label{H_osc_M}
H&=&\int d^3 p \sum_{\lambda} \Big[\left(\omega_\tp+\epsilon_1\frac{m}{\omega_\tp}\right) a^\dagger_{M\lambda}({\bf p}) a_{M\lambda}({\bf p})
+\left(\omega_\tp-\epsilon_1\frac{m}{\omega_\tp}\right) b^\dagger_{M\lambda}({\bf p}) b_{M\lambda}({\bf p})\cr
&+&\left(\frac{\epsilon_5}{2}\,\sgn\,\lambda-i\epsilon_1\frac{\text{p}}{2\omega_\tp}\right)\Big(a_{M\lambda}({\bf p}) a_{M\lambda}(-{\bf p})+b_{M\lambda}({\bf p}) b_{M\lambda}(-{\bf p})\Big)\cr
&-&\left(\frac{\epsilon_5}{2}\,\sgn\,\lambda+i\epsilon_1\frac{\text{p}}{2\omega_\tp}\right)\left(a^\dagger_{M\lambda}({\bf p}) a^\dagger_{M\lambda}(-{\bf p})+b^\dagger_{M\lambda}({\bf p}) b^\dagger_{M\lambda}(-{\bf p})\right)\Big].
\end{eqnarray}
In this form, the $a_M$ and $b_M$-type operators are disentangled and we can diagonalize each set separately. Incidentally, in the BCS language the expression 
\begin{equation}\label{gap}
\delta_ \tp = \frac{\epsilon_5}{2}\,\sgn\,\lambda+i\epsilon_1\frac{\text{p}}{2\Omega_\tp}
%= \frac{1}{2}\sqrt{\epsilon_5^2+\epsilon_1^2\frac{\tp^2}{\Omega_\tp^2}}\ e^{i\delta_ \tp }, \ \ \text{with}\ \ \ \tan\delta_ \tp =\frac{\epsilon_1\tp}{\epsilon_5\Omega_\tp}
\end{equation}
is the analogue of the {\it gap function} \cite{monograph_cond_mat}.

\subsubsection*{Diagonalization of the Hamiltonian and Bogoliubov transformations}
We diagonalize the Hamiltonian as
\begin{eqnarray}\label{H_osc_diag}
H&=\int d^3 p \sum_{\lambda} \Big[\Omega_\tp^+ A^\dagger_{\lambda}({\bf p}) A_{\lambda}({\bf p})+\Omega_\tp^- B^\dagger_{\lambda}({\bf p}) B_{\lambda}({\bf p})\Big],
\end{eqnarray}
by adopting the following Bogoliubov transformations, suggested by the form of the Hamiltonian \eqref{H_osc_M}:
\begin{eqnarray}\label{Ansatz_1}
A_{\lambda}({\bf p})&=&\alpha_\tp^+a_{M\lambda}({\bf p})+i\beta_\tp^+\,e^{i\delta_ \tp  }\,a^\dagger_{M\lambda}(-{\bf p}),\cr
B_{\lambda}({\bf p})&=&\alpha_\tp^-b_{M\lambda}({\bf p})+i\beta_\tp^-\,e^{i\delta_ \tp  }\,b^\dagger_{M\lambda}(-{\bf p}),
\end{eqnarray}
where $\alpha_\tp^\pm,\ \beta_\tp^\pm$  are complex coefficients and $\delta_ \tp  $ are real. They all depend in principle on the helicity, but we omit the helicity index. The quantities $\Omega_\tp^+$ and $\Omega_\tp^-$ are real, having the meaning of energies to be determined. In order for the new operators to satisfy the canonical anticommutation relations
\begin{eqnarray}\label{ACR_new}
\{A_\lambda({\bf p}),A^\dagger_{\lambda'}({\bf k})\}&=&\delta_{\lambda\lambda'}\delta({\bf p}-{\bf k}),\cr
\{B_\lambda({\bf p}),B^\dagger_{\lambda'}({\bf k})\}&=&\delta_{\lambda\lambda'}\delta({\bf p}-{\bf k}),
\end{eqnarray}
with all the other anticommutators being zero, the coefficients in \eqref{Ansatz_1} have to satisfy the conditions
\begin{eqnarray}\label{square_1_osc}
|\alpha_\tp^+|^2+|\beta_\tp^+|^2=1,\cr
|\alpha_\tp^-|^2+|\beta_\tp^-|^2=1.
\end{eqnarray}
In other words, the conditions \eqref{square_1_osc} insure that the transformations \eqref{Ansatz_1} are canonical. Typically, conditions \eqref{square_1_osc} suggest that the Bogoliubov transformations are rotations in the space of creation and annihilation operators, for which a customary notation \cite{Umezawa-book, Bog-Shirk} is
\begin{eqnarray}\label{BT_sin_cos}
\alpha_\tp^+=\cos\varphi_\tp  ^+,\ \ \ \ \beta_\tp^+=-\sin\varphi_\tp  ^+,\cr
\alpha_\tp^-=\cos\varphi_\tp  ^-,\ \ \ \ \beta_\tp^-=-\sin\varphi_\tp  ^-.
\end{eqnarray}

We shall diagonalize the part of the Hamiltonian depending on $a_M,\ a_M^\dagger$. Introducing the Ansatz \eqref{Ansatz_1} into \eqref{H_osc_diag}, we find
\begin{eqnarray}
&&\int d^3 p \sum_{\lambda} \Omega_\tp^+ A^\dagger_{\lambda}({\bf p}) A_{\lambda}({\bf p})=\int d^3 p \sum_{\lambda} \Omega_\tp^+\Big[|\alpha_\tp^+|^2\,a^\dagger_{M\lambda}({\bf p}) a_{M\lambda}({\bf p})+|\beta_\tp^+|^2 a_{M\lambda}(-{\bf p})a^\dagger_{M\lambda}(-{\bf p})\cr
&&-i(\alpha_\tp^+)^*\beta_\tp^+\,e^{i\delta_ \tp  }\,a^\dagger_{M\lambda}({\bf p}) a^\dagger_{M\lambda}(-{\bf p})-i\alpha_\tp^+(\beta_\tp^+)^*\,e^{-i\delta_ \tp  }a_{M\lambda}({\bf p}) a_{M\lambda}(-{\bf p})\Big].
\end{eqnarray}
Identifying the coefficients with those in \eqref{H_osc_M}, we arrive at the following equations:
\begin{eqnarray}\label{coeff_eq_1}
&&|\alpha_\tp^+|^2-|\beta_\tp^+|^2=\frac{\omega_\tp^2+m\epsilon_1}{\Omega_\tp^+\ \omega_\tp},\cr
&&\Omega_\tp^+(\alpha_\tp^+)^*\beta_\tp^+(\cos\delta_ \tp  +i\sin\delta_ \tp  )=-\epsilon_1\frac{\text{p}}{2\omega_\tp}+i\frac{\epsilon_5}{2}\,\sgn\,\lambda,\cr
&&\Omega_\tp^+\alpha_\tp^+(\beta_\tp^+)^*(\cos\delta_ \tp  -i\sin\delta_ \tp  )=-\epsilon_1\frac{\text{p}}{2\omega_\tp}-i\frac{\epsilon_5}{2}\,\sgn\,\lambda.
\end{eqnarray}
From the last two relations in \eqref{coeff_eq_1} we infer that $\alpha_\tp^+$ and $\beta_\tp^+$ can be taken to be real, leading to
\begin{eqnarray*}
&&\alpha_\tp^+\beta_\tp^+\cos\delta_ \tp  =-\epsilon_1\frac{\text{p}}{2\omega_\tp\Omega_\tp^+},\cr
&&\alpha_\tp^+\beta_\tp^+\sin\delta_ \tp  =\frac{\epsilon_5}{2\Omega_\tp^+}\,\sgn\,\lambda.
\end{eqnarray*}
Thus, we obtain
\begin{eqnarray}\label{tg_delta}
\tan\delta_ \tp  =-\sgn\,\lambda\frac{\epsilon_5\omega_\tp}{\epsilon_1\tp},
\end{eqnarray}
from where
\begin{eqnarray}\label{sc_delta}
\sin\delta_ \tp  &=&\frac{\tan\delta_ \tp  }{\sqrt{1+\tan^2\delta_ \tp  }}=-\sgn\,\lambda\frac{\epsilon_5}{\sqrt{\epsilon_5^2+\epsilon_1^2\frac{\tp^2}{\omega_\tp^2}}},\cr
\cos\delta_ \tp  &=&\frac{1}{\sqrt{1+\tan^2\delta_ \tp  }}=\epsilon_1\frac{\tp}{\omega_\tp}\frac{1}{\sqrt{\epsilon_5^2+\epsilon_1^2\frac{\tp^2}{\omega_\tp^2}}}.
\end{eqnarray}
With these results we return to \eqref{coeff_eq_1} and find
\begin{eqnarray}\label{coeff_eq_2}
&&\alpha_\tp^+=-\frac{1}{2\Omega_\tp^+\beta_\tp^+}\sqrt{\epsilon_5^2+\epsilon_1^2\frac{\tp^2}{\omega_\tp^2}},
\end{eqnarray}
which we insert into the first equation of \eqref{coeff_eq_1}:
\begin{eqnarray}\label{coeff_eq_3}
\frac{1}{4(\Omega_\tp^+)^2(\beta_\tp^+)^2}\left(\epsilon_5^2+\epsilon_1^2\frac{\tp^2}{\omega_\tp^2}\right)-(\beta_\tp^+)^2=\frac{\omega_\tp^2+m\epsilon_1}{\Omega_\tp^+\ \omega_\tp}.
\end{eqnarray}
Eqs. \eqref{coeff_eq_2} and \eqref{coeff_eq_3}, together with the requirement of canonicity \eqref{square_1_osc},  are satisfied by the real expressions
\begin{eqnarray}\label{BT_coeff+}
\alpha_\tp^+&=&\sqrt{\frac{\Omega_\tp^++\omega_\tp}{2\Omega_\tp^+}+\frac{m\epsilon_1}{2\omega_\tp\Omega_\tp^+}},\cr
\beta_\tp^+&=&-\sqrt{\frac{\Omega_\tp^+-\omega_\tp}{2\Omega_\tp^+}-\frac{m\epsilon_1}{2\omega_\tp\Omega_\tp^+}},
\end{eqnarray}
where 
\begin{eqnarray}\label{Omega+}
\Omega_\tp^+&=&\sqrt{{\tp}^2+M_+^2},\ \ \ \text{with}\ \ M_+^2=(m+\epsilon_1)^2+\epsilon_5^2.
\end{eqnarray}

Inspecting the Hamiltonian \eqref{H_osc_M}, we notice that the part depending on $b_M,\ b_M^\dagger$ is identical to the part depending on $a_M,\ a_M^\dagger$, up to the substitution $\epsilon_1\to-\epsilon_1$. As a result, we infer immediately the form of the corresponding coefficients:
\begin{eqnarray}\label{BT_coeff-}
\alpha_\tp^-&=&\sqrt{\frac{\Omega_\tp^-+\omega_\tp}{2\Omega_\tp^-}-\frac{m\epsilon_1}{2\omega_\tp\Omega_\tp^-}},\cr
\beta_\tp^-&=&-\sqrt{\frac{\Omega_\tp^--\omega_\tp}{2\Omega_\tp^-}+\frac{m\epsilon_1}{2\omega_\tp\Omega_\tp^-}},
\end{eqnarray}
where 
\begin{eqnarray}\label{Omega-}
\Omega_\tp^-&=&\sqrt{{\tp}^2+M_-^2},\ \ \ \text{with}\ \ M_-^2=(m-\epsilon_1)^2+\epsilon_5^2.
\end{eqnarray}

\subsubsection*{The physical vacuum and the Fock space of the quasiparticles}

The set of operators which diagonalize the Hamiltonian act on a new vacuum $|\Phi_0\rangle $, which satisfies
\begin{equation}\label{AB_vacuum}
A_\lambda({\bf p})|\Phi_0\rangle =B_\lambda({\bf p})|\Phi_0\rangle =0,
\end{equation}
and represents the physical vacuum of the model. The physical particle states are Bogoliubov quasiparticles, of Majorana type, with the definite masses $M_\pm^2=(m\pm\epsilon_1)^2+\epsilon_5^2$.

The relation between $|\Phi_0\rangle $  and the bare particles' vacuum $|0\rangle$ is derived by assuming that the vacuum of quasiparticles is written as an arbitrary superposition of pairs of Majorana particles associated with the fields $\psi_\pm(x)$:
\begin{eqnarray}
|\Phi_0\rangle ={\cal N}\ \Pi_{{\bf p},\lambda}\ e^{R_p^+\,a_{M\lambda}^\dagger({\bf p})a_{M\lambda}^\dagger(-{\bf p})}e^{R_p^-\,b_{M\lambda}^\dagger({\bf p})b_{M\lambda}^\dagger(-{\bf p})}|0\rangle,
\end{eqnarray}
where $\cal N$ is a normalization constant. Using \eqref{AB_vacuum} and \eqref{Ansatz_1}, one finds that $R_p^\pm=-i\beta^\pm_p e^{i\delta_ \tp }/\alpha^\pm_p$. Pauli's principle implies that 
$$
(a_{M\lambda}^\dagger({\bf p})a_{M\lambda}^\dagger(-{\bf p}))^n=(b_{M\lambda}^\dagger({\bf p})b_{M\lambda}^\dagger(-{\bf p}))^n=0, \ \ \ \text{for}\ \ \ \ n>1,
$$
therefore 
\begin{eqnarray}
|\Phi_0\rangle ={\cal N}\ \Pi_{{\bf p},\lambda}\ \left(1+R_p^+\,a_{M\lambda}^\dagger({\bf p})a_{M\lambda}^\dagger(-{\bf p})\right)\left(1+R_p^-\,b_{M\lambda}^\dagger({\bf p})b_{M\lambda}^\dagger(-{\bf p})\right)|0\rangle.
\end{eqnarray}
Recalling \eqref{square_1_osc}, we note that
\begin{eqnarray*}
&&\langle 0|\Big(\alpha_\tp^+ +i\beta^+_p e^{-i\delta_ \tp }\,a_{M\lambda}({\bf p})a_{M\lambda}(-{\bf p})\Big)\left(\alpha_\tp^+ -i\beta^+_p e^{i\delta_ \tp }\,a_{M\lambda}^\dagger({\bf p})a_{M\lambda}^\dagger(-{\bf p})\right)|0\rangle\cr
&=&
\langle 0|\Big(|\alpha_\tp^+|^2+|\beta_\tp^+|^2\,a_{M\lambda}({\bf p})a_{M\lambda}({\bf p})^\dagger a_{M\lambda}(-{\bf p})a_{M\lambda}^\dagger(-{\bf p})\Big)|0\rangle\cr
&=&\langle 0|\Big(|\alpha_\tp^+|^2+|\beta_\tp^+|^2\,(1+a_{M\lambda}^\dagger({\bf p})a_{M\lambda}({\bf p}))(1+ a_{M\lambda}^\dagger(-{\bf p})a_{M\lambda}(-{\bf p}))\Big)|0\rangle
\cr
&=&\langle 0|\Big(|\alpha_\tp^+|^2+|\beta_\tp^+|^2\Big)|0\rangle=1,
\end{eqnarray*}
leading to the normalized quasiparticle vacuum in the form:
\begin{eqnarray}\label{normalized vacuum}
|\Phi_0\rangle =\ \Pi_{{\bf p},\lambda}\  \left(\alpha_\tp^+ -i\beta_\tp^+e^{i\delta_ \tp }\,a_{M\lambda}^\dagger({\bf p})a_{M\lambda}^\dagger(-{\bf p})\right)\left(\alpha_\tp^- -i\beta_\tp^-e^{i\delta_ \tp }\,b_{M\lambda}^\dagger({\bf p})b_{M\lambda}^\dagger(-{\bf p})\right)|0\rangle.
\end{eqnarray}
Just as in the BCS theory, the phase of the "Cooper pairs" of bare Majorana particles is given by the phase of the gap function \eqref{gap}. This phase, in the present case, is fixed by the choice of the parameters $m,\epsilon_1,\epsilon_5$ in the Lagrangian \eqref{Lagr_osc} and for each pair depends on the momentum of its constituents. The physical vacuum is therefore unique \footnote{In constrast, in the BCS theory or NJL model, the phase of the gap function is arbitrary due to the $U(1)$ symmetry of the Lagrangian, and its variation leads to an infinity of degenerate vacua, what is the essence of spontaneous breaking of symmetry.}.

The bare Majorana particles composing the pairs have opposite momenta and spins, consistent with the Poincar\'e invariance which implies energy-momentum and angular momentum conservation. 

The Fock space built on the vacuum $|\Phi_0\rangle $ consists of Majorana particle states with two different masses, $M_+$ and $M_-$ given by \eqref{Omega+} and \eqref{Omega-}:
\begin{eqnarray}\label{H_on_quasipart}
H\,A_\lambda^\dagger({\bf p})|\Phi_0\rangle &=&\Omega_\tp^+A_\lambda^\dagger({\bf p})|\Phi_0\rangle ,\cr
H\,B_\lambda^\dagger({\bf p})|\Phi_0\rangle &=&\Omega_\tp^-B_\lambda^\dagger({\bf p})|\Phi_0\rangle .
\end{eqnarray}
These quasiparticles, with indefinite baryon number, are the only physical particles in the model. Neutron and antineutron do not exist as particle states.

\subsubsection*{Vacuum condensate and baryon-number violation}

Coleman's theorem states that "the invariance of the vacuum is the invariance of the world" \cite{Coleman}.
We therefore expect to see violation of baryonic number in the vacuum condensate. As mentioned earlier, the bare neutron and antineutron states have definite baryonic numbers. On the other hand, the baryonic number is undefined for the states of bare Majorana particles, $a_{M\lambda}^\dagger({\bf p})|0\rangle$ and $b_{M\lambda}^\dagger({\bf p})|0\rangle$. We may attempt to rewrite the vacuum condensate as superposition of pairs of bare neutrons and antineutrons, $a_{\lambda}^\dagger({\bf p})a_{\lambda}^\dagger(-{\bf p})|0\rangle$ and $b_{\lambda}^\dagger({\bf p})b_{\lambda}^\dagger(-{\bf p})|0\rangle$. To this end, we insert  \eqref{M-D} into the Bogoliubov transformations \eqref{Ansatz_1} and find:
\begin{eqnarray}\label{BT_ab}
A_{\lambda}({\bf p})&=&\frac{1}{\sqrt2}\left(\alpha_\tp^+a_{\lambda}({\bf p})+i\beta_\tp^+\,e^{i\delta_ \tp  }\,a^\dagger_{\lambda}(-{\bf p})\right)+\frac{\sgn\lambda}{\sqrt2}\left(\alpha_\tp^+b_{\lambda}({\bf p})+i\beta_\tp^+\,e^{i\delta_ \tp  }\,b^\dagger_{\lambda}(-{\bf p})\right),\cr
B_{\lambda}({\bf p})&=&\frac{1}{\sqrt2}\left(\alpha_\tp^-a_{\lambda}({\bf p})+i\beta_\tp^-\,e^{i\delta_ \tp  }\,a^\dagger_{\lambda}(-{\bf p})\right)-\frac{\sgn\lambda}{\sqrt2}\left(\alpha_\tp^-b_{\lambda}({\bf p})+i\beta_\tp^-\,e^{i\delta_ \tp  }\,b^\dagger_{\lambda}(-{\bf p})\right).
\end{eqnarray}
The requirement on the physical vacuum \eqref{AB_vacuum} then implies:
\begin{eqnarray}\label{vac_ab_A}
\left(\alpha_\tp^+a_{\lambda}({\bf p})+i\beta_\tp^+\,e^{i\delta_ \tp  }\,a^\dagger_{\lambda}(-{\bf p})\right)|\Phi_0\rangle =0,\cr
\left(\alpha_\tp^+b_{\lambda}({\bf p})+i\beta_\tp^+\,e^{i\delta_ \tp  }\,b^\dagger_{\lambda}(-{\bf p})\right)|\Phi_0\rangle =0,\cr
\end{eqnarray}
simultaneously with 
\begin{eqnarray}\label{vac_ab_B}
\left(\alpha_\tp^-a_{\lambda}({\bf p})+i\beta_\tp^-\,e^{i\delta_ \tp  }\,a^\dagger_{\lambda}(-{\bf p})\right)|\Phi_0\rangle =0,\cr
\left(\alpha_\tp^-b_{\lambda}({\bf p})+i\beta_\tp^-\,e^{i\delta_ \tp  }\,b^\dagger_{\lambda}(-{\bf p})\right)|\Phi_0\rangle =0.
\end{eqnarray}
As long as $\alpha_\tp^+\neq\alpha_\tp^-$ and $\beta_\tp^+\neq\beta_\tp^-$, the relations \eqref{vac_ab_A} and \eqref{vac_ab_B} are in conflict. Consequently, for the general case with arbitrary $\epsilon_1$ and $\epsilon_5$ parameters, we have to content ourselves with the expression \eqref{normalized vacuum} for the vacuum condensate.

In the specific case when $\epsilon_1=0$, we notice that
\begin{eqnarray}\label{coeff_BT_e5}
&&M_\pm=M=\sqrt{m^2+\epsilon_5^2},\cr
&&\alpha_\tp^\pm=\alpha_\tp=\sqrt{\frac{\Omega_\tp+\omega_\tp}{2\Omega_\tp}},\cr
&&\beta_\tp^\pm=\beta_\tp=-\sqrt{\frac{\Omega_\tp-\omega_\tp}{2\Omega_\tp}},\cr
&&\sin\delta_ \tp =-\sgn\lambda,\ \ \ \cos\delta_ \tp =0.
\end{eqnarray}
This is the only instance when the relations \eqref{vac_ab_A} and \eqref{vac_ab_B} are compatible and the physical vacuum can be written as
\begin{eqnarray}\label{vacuum_ab}
|\Phi_0\rangle |_{\epsilon_1=0}=\ \Pi_{{\bf p},\lambda}\  \left(\alpha_\tp -\sgn\lambda\,\beta_\tp\,a_{\lambda}^\dagger({\bf p})a_{\lambda}^\dagger(-{\bf p})\right)\left(\alpha_\tp -\sgn_\lambda\,\beta_\tp\,b_{\lambda}^\dagger({\bf p})b_{\lambda}^\dagger(-{\bf p})\right)|0\rangle,
\end{eqnarray}
with the pairs or bare neutrons and antineutrons carrying baryon number $\pm 2$, and thus explicitly exhibiting the baryon-number violation\footnote{Incidentally, if $\epsilon_1,\epsilon_5\ll m$ and we expand the coefficients of the general Bogoliubov transformation \eqref{Ansatz_1} to second order in $\epsilon_1$ and $\epsilon_5$ (see \eqref{approx} below), we find again $\alpha_\tp^+=\alpha_\tp^-$ and $\beta_\tp^+=\beta_\tp^-$, and the vacuum condensate can be recast in a form similar to \eqref{vacuum_ab}. In Ref. \cite{chang}, for example, the physical vacuum was derived in this approximation, for $\epsilon_5=0$.}.

\subsubsection*{Unitary inequivalence of representations}

Let us calculate the inner product of the two vacua, using  \eqref{normalized vacuum} and taking into account \eqref{naive_vac}, \eqref{BT_coeff+} and \eqref{BT_coeff-}: 
\begin{eqnarray}
\langle 0|\Phi_0\rangle &=& \Pi_{{\bf p},\lambda}\ |\alpha_\tp^+||\alpha_\tp^-|=\Pi_{{\bf p},\lambda}\ \frac{1}{2}\left(1+\frac{\omega_\tp}{\Omega^+_p}+\frac{m\epsilon_1}{\omega_\tp\Omega_\tp^+}\right)^{1/2}\left(1+\frac{\omega_\tp}{\Omega^-_p}-\frac{m\epsilon_1}{\omega_\tp\Omega_\tp^-}\right)^{1/2}\\
&=&\exp\left(\int\frac{d^3p}{(2\pi)^3}\sum_\lambda\frac{1}{2} \ln \left[\frac{1}{2}\left(1+\frac{\omega_\tp}{\Omega^+_p}+\frac{m\epsilon_1}{\omega_\tp\Omega_\tp^+}\right)\right]\left[\frac{1}{2}\left(1+\frac{\omega_\tp}{\Omega^-_p}-\frac{m\epsilon_1}{\omega_\tp\Omega_\tp^-}\right)\right]\right).
\nonumber
\end{eqnarray}
In the large momentum limit,  $\left[\frac{1}{2}\left(1+\frac{\omega_\tp}{\Omega^+_p}+\frac{m\epsilon_1}{\omega_\tp\Omega_\tp^+}\right)\right]\left[\frac{1}{2}\left(1+\frac{\omega_\tp}{\Omega^-_p}-\frac{m\epsilon_1}{\omega_\tp\Omega_\tp^-}\right)\right]\approx 1-\frac{\epsilon_1^2+\epsilon_5^2}{2\tp^2}$,  and the exponential diverges as $\exp\left[-(\epsilon_1^2+\epsilon_5^2)\int d{ \tp}\right]$, which leads to the orthogonality of the two vacua,
\begin{equation}
\langle 0|\Phi_0\rangle =0.
\end{equation}

The Fock spaces built on the bare vacuum $|0\rangle$ and on the quasiparticle vacuum $|\Phi_0\rangle $ are, consequently, also orthogonal. (This can be easily confirmed by taking the inner product of two arbitrary states belonging to the two spaces.) The latter is the physical one, while the former is an auxiliary space, an artifact of the quantization method. Although the bare particle states cannot be found among the physical states, this does not mean that the bare operators cannot act on the physical vacuum. The operators $a,a^\dagger,b,b^\dagger$ act on $|\Phi_0\rangle $ through their relations to the quasiparticle operators, i.e. the inverse Bogoliubov transformations \eqref{inverse_BT} together with \eqref{D-M}, always creating and annihilating particles with masses $M_\pm$ and never bare particles of mass $m$. This feature will be used further in defining neutron and antineutron "states" in Sect. \ref{trans_prob}.

\subsection{Heisenberg fields}

Having diagonalized the Hamiltonian \eqref{H_osc_M} in the Schr\"odinger picture, we can now easily move to the Heisenberg picture. We have obtained the solutions of the Hamiltonian \eqref{Hamilt_osc} as two non-degenerate Majorana fields. Their time evolution is given by
\begin{equation}
e^{iHt}\Psi_\pm({\bf x},0)e^{-iHt}=\Psi_\pm({\bf x},t).
\end{equation}
The corresponding creation and annihilation operators evolve as
\begin{eqnarray}
A({\bf p},t)&=&e^{iHt}A({\bf p})e^{-iHt}=A({\bf p})e^{-i\Omega_\tp^{+}t},\cr
A^\dagger({\bf p},t)&=&e^{iHt}A^\dagger({\bf p}) e^{-iHt}=A^\dagger({\bf p})e^{i\Omega_\tp^{+}t},\cr
B({\bf p},t)&=&e^{iHt}B({\bf p})e^{-iHt}=B({\bf p})e^{-i\Omega_\tp^{+}t},\cr
B^\dagger({\bf p},t)&=&e^{iHt}B^\dagger({\bf p}) e^{-iHt}=B^\dagger({\bf p})e^{i\Omega_\tp^{+}t},
\end{eqnarray}
where we used $H$ in the form \eqref{H_osc_diag}.
Thus, the primary time-dependent Majorana fields will read:
\begin{eqnarray}\label{Majorana_mode_exp_H}
\Psi_+({\bf x},t)=\int\frac{d^3p}{(2\pi)^{3/2}\sqrt{2\Omega_\tp^+}}\sum_{\lambda}\left(A_\lambda({\bf p})U_\lambda({\bf p})e^{-i(\Omega_\tp^+ t-{\bf p\cdot x })}+\sgn \lambda\, A^\dagger_\lambda({\bf p})V_\lambda({\bf p})e^{i(\Omega_\tp^+ t-{\bf p\cdot x })}\right),\cr
\Psi_-({\bf x},t)=\int\frac{d^3p}{(2\pi)^{3/2}\sqrt{2\Omega_\tp^-}}\sum_{\lambda}\left(B_\lambda({\bf p})\tilde U_\lambda({\bf p})e^{-i(\Omega_\tp^- t-{\bf p\cdot x })}-\sgn \lambda\, B^\dagger_\lambda({\bf p})\tilde V_\lambda({\bf p})e^{i(\Omega_\tp^- t-{\bf p\cdot x })}\right),
\end{eqnarray}
with the spinors $U_\lambda({\bf p}), V_\lambda({\bf p})$ and $\tilde U_\lambda({\bf p}),\tilde V_\lambda({\bf p})$ satisfying the equations  of motion \eqref{EOM_gen} in momentum space.

Inverting the Bogoliubov transformations \eqref{Ansatz_1}, namely
\begin{eqnarray}\label{inverse_BT}
a_{M\lambda}({\bf p})=\alpha_\tp^+A_{\lambda}({\bf p})-i\beta_\tp^+\,e^{i\delta_ \tp  }\,A^\dagger_{\lambda}(-{\bf p}),\cr
b_{M\lambda}({\bf p})=\alpha_\tp^-B_{\lambda}({\bf p})-i\beta_\tp^-\,e^{i\delta_ \tp  }\,B^\dagger_{\lambda}(-{\bf p}),
\end{eqnarray}
we obtain the time evolution of the operators $a_{M\lambda}({\bf p}), b_{M\lambda}({\bf p})$:
\begin{eqnarray}\label{inverse_BT_t}
a_{M\lambda}({\bf p},t)=\alpha_\tp^+A_{\lambda}({\bf p})e^{-i\Omega_\tp^{+}t}-i\beta_\tp^+\,e^{i\delta_ \tp  }\,A^\dagger_{\lambda}(-{\bf p})e^{i\Omega_\tp^{+}t},\cr
b_{M\lambda}({\bf p},t)=\alpha_\tp^-B_{\lambda}({\bf p})e^{-i\Omega_\tp^{-}t}-i\beta_\tp^-\,e^{i\delta_ \tp  }\,B^\dagger_{\lambda}(-{\bf p})e^{i\Omega_\tp^{-}t}.
\end{eqnarray}
Using \eqref{D-M} and \eqref{inverse_BT_t}, we can express the time-dependent $a_{\lambda}({\bf p},t), b_{\lambda}({\bf p},t)$ as well.

\subsection{Diagonalization of Hamiltonian and primary Majorana fields}\label{primary Majoranas}

In the typical cases of mass shift of Dirac fermions by vacuum condensate encountered in the NJL model \cite{NJL}, \cite{UTK}, the Bogoliubov transformations relating the creation and annihilation operators of different masses can be obtained by two equivalent procedures. One of them is what we have described above: having derived the Hamiltonian of the system, $H=H_0+H_{int}$, the field $\Psi_D(x)$ is replaced in the Hamiltonian, at $t=0$ (Schr\"odinger picture), by the solution $\psi_D(x)$ of the equation of motion $i\partial_t\psi_D(x)=[\psi_D(x),H_0]$ , i.e.
\begin{equation}\label{NJL_D}
\Psi_D({\bf x},0)=\psi_D({\bf x},0),
\end{equation}
and the Hamiltonian is subsequently diagonalized by using Bogoliubov transformations. The quasiparticle operators that diagonalize the total Hamiltonian will be the creation and annihilation operators of the field $\Psi_D(x)$, which satisfies the equation of motion $i\partial_t\Psi_D(x)=[\Psi_D(x),H]$. Typically, the bare field  $\psi_D(x)$ and the quasiparticle field $\Psi_D(x)$ are free Dirac fields with different masses. In this way, one finds the solution  $\Psi_D(x)$ without solving directly its equation of motion. This method is essentially a relativistic extension of Bogoliubov's approach to the theories of superfluidity and superconductivity \cite{Bogoliubov}.

The second procedure is the one used in the work of Nambu and Jona-Lasinio \cite{NJL}: knowing the Hamiltonian $H$, one solves the equation of motion $i\partial_t\Psi_D(x)=[\Psi_D(x),H]$, and subsequently identifies its solution, at $t=0$, with the solution of $i\partial_t\psi_D(x)=[\psi_D(x),H_0]$. In other words, one imposes the "boundary condition" \eqref{NJL_D} to the two known solutions. In this case, the purpose is strictly to find the Bogoliubov transformations and the relation between the bare particle vacuum and the quasiparticle vacuum. The results are the same as those obtained by the Hamiltonian diagonalization method.

The mixing of fields in the baryon-number violating model that we have been analyzing requires more care in the application of the procedures outlined above. We have seen that the Hamiltonian diagonalization procedure succeeds when using the identification \eqref{NJL},
\begin{equation*}
\Psi({\bf x},0)=\psi({\bf x},0).
\end{equation*}
Recall that the field $\psi(x)$ is a Dirac field of mass $m$, while $\Psi(x)$ is not a Dirac, nor a Majorana, field. In effect, the field $\Psi(x)$ does not satisfy a simple equation of motion, but an equation in which it is mixed with its charge conjugate $\Psi^c(x)$, eq. \eqref{eom_Psi}.  A "rotation" of the creation and annihilation operators of $\psi(x)$ does not take us to new creation and annihilation operators, because there are no such operators for the field $\Psi(x)$. This is an indication that the second procedure outlined above cannot work with the boundary condition \eqref{NJL}.

In hindsight, we realize that the actual identification of fields for which \eqref{NJL} was standing was
\begin{equation}\label{NJL_Majorana}
\Psi_\pm({\bf x},0)=\psi_\pm({\bf x},0),
\end{equation}
where $\psi_\pm(x)$ satisfy
\begin{eqnarray*}
(i\gamma^{\mu}\partial_{\mu}-m)\psi_\pm(x)=0
\end{eqnarray*}
and $\Psi_\pm(x)$ satisfy eqs. \eqref{EOM_gen},
\begin{eqnarray*}
(i\gamma^{\mu}\partial_{\mu}-(m+\epsilon_1)-i\epsilon_5\gamma_5)\Psi_+(x)=0,\cr
(i\gamma^{\mu}\partial_{\mu}-(m-\epsilon_1)+i\epsilon_5\gamma_5) \Psi_-(x)=0.
\end{eqnarray*}
We call the fields $\Psi_\pm$ {\it primary Majorana fields}, as they are the simplest combinations of the neutron field $\Psi$ and its charge conjugate $\Psi^c$,  which satisfy uncoupled equations of motion. The two non-degenerate primary Majorana fields can be related to the two mass-degenerate bare Majorana fields by different "rotations" of the creation and annihilation operators. In Appendix  \ref{NJL_calculation} we shall prove that the Bogoliubov transformations \eqref{Ansatz_1}, with the coefficients specified by \eqref{BT_coeff+} and \eqref{BT_coeff-}, can be found also by the second procedure outlined above, starting from the boundary condition \eqref{NJL_Majorana}.

We emphasize specifically the r\^ole of the primary Majorana fields, because in certain situations one can choose other combinations of Majorana fields that diagonalize the Lagrangian as well. For example, when $\epsilon_1=0,\epsilon_5\neq 0$, the Lagrangian is diagonal in terms of $\Psi_\pm(x)$, but also in terms of the Dirac-type fields $N(x)$ which satisfy the Dirac equation \eqref{Dirac_N} (see the discussion in Sect. \ref{propagator_RBT}) and are related to $\Psi$ and $\Psi^c$ by the relativistic Bogoliubov transfomation \eqref{RBT_def}. Due to the simplicity of the equation of motion for $N(x)$, it may be tempting to use the relativistic Bogoliubov transformation as the basis for the boundary condition at $t=0$, namely to make the identification
\begin{eqnarray}
\left(\begin{array}{c}
            \psi({\bf x},0)\\
            \psi^{c}({\bf x},0)
            \end{array}\right)=\left(\begin{array}{c}
            \Psi({\bf x},0)\\
            \Psi^{c}({\bf x},0)
            \end{array}\right)
&=& \left(\begin{array}{c}
            \cos\Theta N({\bf x},0)-i\gamma_{5}\sin\Theta N^c({\bf x},0)\\
            \cos\Theta N^c({\bf x},0)-i\gamma_{5}\sin\Theta N({\bf x},0)
            \end{array}\right).
\end{eqnarray}
In this case, the resulting transformations between the operators of $\psi(x)$ and those of $N(x)$ are essentially incompatible, in the sense that the two annihilation operators of $N(x)$, say $A_{N\lambda}({\bf p})$ and $B_{N\lambda}({\bf p})$, do not destroy the same vacuum condensate $|\Phi_{N0}\rangle$. This inconsistency does not appear if one adheres to primary fields and formula \eqref{NJL_Majorana}.

The identification of primary fields is an essential step in treating any quantum systems with mixings of fields, like the seesaw mechanism Lagrangian or various models of neutrino mixing and oscillation. 

%%%%%%%%%%%%%%%%%%%%%%%%%%%%%%%%%%5

\section{Neutron states in physical Fock space and the $n-\bar n$ transition probability}\label{trans_prob}

When we embed the quadratic Lagrangian \eqref{Lagr_osc} into the Standard Model, the field $\Psi(x)$ plays the role of neutron field and takes part in the neutron interactions already present there. At the same time, we have to give up the picture of the neutron as a particle with definite mass and flavour. It is then necessary to redefine the notion of neutron and antineutron, when there are no creation and annihilation operators for them.

The only possibility for a consistent definition is to associate the neutron and antineutron with the field $\Psi(x)$, in other words, to define these "states" by their dynamical relations with the other particles with which they interact.  The natural procedure is  to use the Schr\"odinger picture identification \eqref{NJL},
$$\Psi({\bf x},0)=\psi({\bf x},0),$$ 
together with the consistency requirement that, in the limit when the baryon-number violating interaction vanishes (i.e. $\epsilon_1,\epsilon_5\to0$), one recovers the bare, or flavour, neutron state defined on the vacuum $|0\rangle$. In practice, we start by Fourier transforming the field $\psi({\bf x},0)$:
$$
\int\frac{d^3x}{(2\pi)^{3/2}}e^{i\px}\bar\psi({\bf x},0) =\frac{1}{\sqrt{2\omega_\tp}}\sum_{\lambda'}\left(a_{\lambda'}^\dagger({\bf p})\bar u_{\lambda'}({\bf p})+b_{\lambda'}(-{\bf p})\bar v_{\lambda}(-{\bf p})\right).
$$
Upon multiplication by $\gamma_0u_\lambda({\bf p})$ and use of the relations \eqref{spinor_norm}, we find
\begin{equation}\label{bare_neutron_creation}
\frac{1}{\sqrt{2\omega_\tp}}\left(\int\frac{d^3x}{(2\pi)^{3/2}}e^{i\px}\bar\psi({\bf x},0) \right)\gamma_0u_\lambda({\bf p})=a_{\lambda}^\dagger({\bf p}).
\end{equation}
The operator in the left-hand side of \eqref{bare_neutron_creation}, acting on the vacuum $|0\rangle$, produces the bare neutron state $a_{\lambda}^\dagger({\bf p})|0\rangle$. We shall therefore adopt it as the definition of the "neutron creation operator" on the physical vacuum\footnote{In Ref. \cite{chang}, the definition of the neutron state is (with our notations and conventions) $\sqrt{\frac{\omega_\tp}{2}}\frac{1}{m}\left(\int\frac{d^3x}{(2\pi)^{3/2}}e^{i\px}\bar\psi({\bf x},0) \right)u_\lambda({\bf p})=a_{\lambda}^\dagger({\bf p})+i\frac{\tp}{m}\sgn\,\lambda\ b_{\lambda}(-{\bf p})$. However, this expression does not give sensible results when applied to multiparticle (antineutron) states in the limit when the baryon-number violating interaction vanishes, therefore it cannot be a proper "neutron creation operator" in any setup. Moreover, it is not applicable to the case when bare particles are massless.}, in which case it is preferable to also replace $\psi({\bf x},0)$ by $\Psi({\bf x},0)$:
\begin{eqnarray}\label{n_state}
|n({\bf p},\lambda)\rangle&\equiv&\frac{1}{\sqrt{2\omega_\tp}}\left(\int\frac{d^3x}{(2\pi)^{3/2}}e^{i\px}\bar\Psi({\bf x},0) \right)\gamma_0u_\lambda({\bf p})|\Phi_0\rangle =a_{\lambda}^\dagger({\bf p})|\Phi_0\rangle \cr
&=&\frac{1}{\sqrt 2}\left(\alpha_\tp^+A^\dagger_\lambda({\bf p})+\alpha_\tp^-B^\dagger_\lambda({\bf p})\right)|\Phi_0\rangle,
\end{eqnarray}
with the coefficients given by \eqref{BT_coeff+}--\eqref{Omega-}. In writing the final expression of \eqref{n_state}, we used \eqref{D-M}, \eqref{AB_vacuum} and the inverses of the Bogoliubov transformations \eqref{inverse_BT}. (Alternatively, we would obtain the same expression by direct calculation, following the method used in Appendix \ref{NJL_calculation}.)
Similar considerations for the antineutron state lead to the definition:
\begin{eqnarray}\label{nbar_state}
|\bar n({\bf p},\lambda)\rangle&\equiv&\sgn\,\lambda\frac{1}{\sqrt{2\omega_\tp}}\left(\int\frac{d^3x}{(2\pi)^{3/2}}e^{i\px}\overline{\Psi^c}({\bf x},0)\right)\gamma_0u_\lambda({\bf p})|\Phi_0\rangle =b_{\lambda}^\dagger({\bf p})|\Phi_0\rangle \cr
&=&\frac{1}{\sqrt 2}\sgn\,\lambda\left(\alpha_\tp^+A^\dagger_\lambda({\bf p})-\alpha_\tp^-B^\dagger_\lambda({\bf p})\right)|\Phi_0\rangle.
\end{eqnarray}
We took advantage of the fact that, in spite of the $a$ and $b$-type operators not being creation and annihilation operators on the physical vacuum $|\Phi_0\rangle $, this does not prevent us from defining their action on this vacuum, which is achieved through the inverse Bogoliubov transformations. Thus, neutron and antineutron states are naturally defined on the physical Fock space. 

The oscillation amplitude between neutron and antineutron is obtained by letting the neutron state evolve and sampling the amount of antineutron in it at an arbitrary time $t$:
\begin{equation}\label{trans_1}
A_{n\bar n}=\langle \bar n({\bf p},\lambda)|n({\bf p},\lambda),t\rangle\equiv\langle \bar n({\bf p},\lambda)| e^{-iHt}|n({\bf p},\lambda)\rangle.
\end{equation}
Using \eqref{n_state} and \eqref{nbar_state}, 
as well as the Hamiltonian in the form \eqref{H_osc_diag} and its action on the quasiparticle states \eqref{H_on_quasipart}, we obtain:
\begin{eqnarray}\label{trans_2}
A_{n\bar n}=\frac{1}{2}\,\sgn\,\lambda\Big[(\alpha_\tp^+)^2e^{-i\Omega_\tp^+t}-(\alpha_\tp^-)^2e^{-i\Omega_\tp^-t}\Big],
\end{eqnarray}
with the various coefficients and energies given by \eqref{BT_coeff+}--\eqref{Omega-}. This is the general expression, valid for any values of the parameters $m$, $\epsilon_1$ and $\epsilon_5$ in the Lagrangian \eqref{Lagr_osc}.

To come to a more familiar expresion of the transition amplitude, we shall consider $\epsilon_1,\epsilon_5\ll m$ and expand \eqref{trans_2} to second order in $\epsilon_1$ and $\epsilon_5$. In this order,
\begin{eqnarray}\label{approx}
\Omega_\tp^\pm&=&\sqrt{\omega_\tp^2\pm2m\epsilon_1+\epsilon_1^2+\epsilon_5^2}\approx\omega_\tp\left[1\pm\frac{m\epsilon_1}{\omega_\tp^2}+\frac{1}{2\omega_\tp^2}\left(1-\frac{3m^2}{\omega_\tp^2}\right)\epsilon_1^2+\frac{\epsilon_5^2}{2\omega_\tp^2}\right]\cr&=&\omega_\tp\left[1\pm\frac{m\epsilon_1}{\omega_\tp^2}+\Delta(\epsilon_1^2,\epsilon_5^2)\right],\cr
(\alpha_\tp^\pm)^2&=&\frac{\Omega_\tp^\pm+\omega_\tp}{2\Omega_\tp^\pm}\pm\frac{m\epsilon_1}{2\omega_\tp\Omega_\tp^\pm}=\frac{1}{2}+\frac{\omega_\tp}{2\Omega_\tp^\pm}\left(1\pm\frac{m\epsilon_1}{\omega_\tp^2}\right)\approx 1-\frac{1}{4\omega_\tp^2}\left(\frac{\tp^2}{\omega_\tp^2}\epsilon_1^2+\epsilon_5^2\right).
%
%(\beta_\tp^\pm)^2&=&\frac{\Omega_\tp^\pm-\omega_\tp}{2\Omega_\tp^\pm}\mp\frac{m\epsilon_1}{2\omega_\tp\Omega_\tp^\pm}=\frac{1}{2}-\frac{\omega_\tp}{2\omega_\tp^\pm}\left(1\pm\frac{m\epsilon_1}{\omega_\tp^2}\right)\approx\frac{1}{4\omega_\tp^2}\left(\frac{\tp^2}{\omega_\tp^2}\epsilon_1^2+\epsilon_5^2\right),\cr
%
%\alpha_\tp^\pm\beta_\tp^\pm&\approx&\frac{1}{2\omega_\tp}\left(\frac{\tp^2}{\omega_\tp^2}\epsilon_1^2+\epsilon_5^2\right)^{1/2}.
\end{eqnarray}
Returning with  \eqref{approx} into \eqref{trans_2}, we find the transition amplitude
\begin{eqnarray}\label{trans_approx}
A_{n\bar n}=&-i\,\sgn\,\lambda\,e^{-i\omega_\tp\left(1+\Delta(\epsilon_1^2,\epsilon_5^2)\right)t}\left(1-\frac{1}{4\omega_\tp^2}\left(\frac{\tp^2}{\omega_\tp^2}\epsilon_1^2+\epsilon_5^2\right)\right)\sin\left(\frac{\epsilon_1m}{\omega_\tp}t\right),
\end{eqnarray}
leading to the neutron-antineutron transition probability
\begin{eqnarray}\label{prob_approx}
P_{n\bar n}=&\sin^2\left(\frac{\epsilon_1m}{\omega_\tp}t\right).
\end{eqnarray}

The probability formula \eqref{prob_approx} shows that the neutron-antineutron transition is practically unaffected by the (CP-violation) parameter $\epsilon_5$. The result coincides with the usual free oscillation probability obtained in the framework of quantum mechanics in the same limit, i.e. $\epsilon_1,\epsilon_5\ll m$.

When $\epsilon_1=0$ and $\epsilon_5\neq 0$, the free transition probability vanishes exactly, as can be easily seen from \eqref{trans_2}, taking into account that in this case $\alpha_\tp^+=\alpha_\tp^-$ and $\Omega_\tp^+=\Omega_\tp^-$  (see \eqref{coeff_BT_e5}). This is of course expected, because the two primary Majorana fields are in this case degenerate in mass. However, as we shall see in the next section, the anomalous baryon-number violating propagator will still be nonvanishing \cite{FT1,FT3}.

%%%%%%%%%%%%%%%%%%%%%%%%%%%%%%%%%%%%%%%%%

\section{Anomalous propagator}\label{propagator}

The purpose of this section is to show that  the canonical quantization procedure outlined in Sect. \ref{Hamilt_formalism} and \ref{trans_prob} is compatible with the results obtained in the Lagrangian/path integral approach. This comparison will give more support to the definition which we adopted for the neutron and antineutron states. 
Ordinarily, by the token of fermion number conservation, in perturbative Standard Model, we expect the propagator $\Psi_\pm(x)$
\begin{equation}\label{anom_prop}
\langle T\Psi^c(x)\bar\Psi(y)\rangle=\theta(x_0-y_0)\langle \Psi^c(x)\bar\Psi(y)\rangle-\theta(y_0-x_0)\langle \bar\Psi(y)\Psi^c(x)\rangle
\end{equation}
to vanish. However, since baryon number conservation is now violated, the above propagator is nonzero. The calculation of this propagator in the canonical framework described above and by means of the manifestely relativistic Lagrangian formalism/path integral will show the coincidence of the two approaches. 

For our purpose, it is sufficient to calculate, in the Hamiltonian formulation, the transition amplitude $\langle \Phi_0|\Psi^c({\bf x},t)\bar\Psi({\bf y},0)|\Phi_0\rangle$ and compare it with the result of the path integral approach.

Using \eqref{NJL}, \eqref{Dirac_mode_exp} and \eqref{charge_conj}, we find:
\begin{eqnarray}\label{nonrel-prop}
&&\langle \Phi_0|\Psi^c({\bf x},t)\bar\Psi({\bf y},0)|\Phi_0\rangle=\int\frac{d^3p\,d^3k}{(2\pi)^3 2(\omega_\tp\omega_k)^{1/2}}\sum_{\lambda,\lambda'}\sgn\lambda\cr
&&\times\langle \Phi_0|\Big[a^\dagger_\lambda({\bf p},t)a^\dagger_{\lambda'}({\bf k})v_\lambda({\bf p})\bar u_{\lambda'}({\bf k})e^{-i({\bf p\cdot x }+{\bf k\cdot y })}+
b_\lambda({\bf p},t)b_{\lambda'}({\bf k})u_\lambda({\bf p})\bar v_{\lambda'}({\bf k})e^{i({\bf p\cdot x }+{\bf k\cdot y })}\cr
&&+a^\dagger_\lambda({\bf p},t)b_{\lambda'}({\bf k})v_\lambda({\bf p})\bar v_{\lambda'}({\bf k})e^{-i({\bf p\cdot x }-{\bf k\cdot y })}
+b_\lambda({\bf p},t)a^\dagger_{\lambda'}({\bf k})u_\lambda({\bf p})\bar u_{\lambda'}({\bf k})e^{i({\bf p\cdot x }-{\bf k\cdot y })}
\Big]|\Phi_0\rangle .
\end{eqnarray}
With the help of \eqref{D-M} and \eqref{inverse_BT_t}, we obtain the vacuum values involved in \eqref{nonrel-prop}:
\begin{eqnarray}\label{vv}
&&\langle \Phi_0|a^\dagger_\lambda({\bf p},t)a^\dagger_{\lambda'}({\bf k})|\Phi_0\rangle =\frac{i}{2}\left(\alpha_\tk ^+\beta_\tp^+e^{-i\delta_ \tp  }e^{-i\Omega_\tp^+t}+\alpha_\tk ^-\beta_\tp^-e^{-i\delta_ \tp  }e^{-i\Omega_\tp^-t}\right)\delta_{\lambda\lambda'}\delta({\bf p}+{\bf k}),\cr
&&\langle \Phi_0|b_\lambda({\bf p},t)b_{\lambda'}({\bf k})|\Phi_0\rangle =-\frac{i}{2}\left(\alpha_\tp^+\beta_\tk ^+e^{i\delta_\tk }e^{-i\Omega_\tp^+t}+\alpha_\tp^-\beta_\tk ^-e^{i\delta_\tk }e^{-i\Omega_\tp^-t}\right)\delta_{\lambda\lambda'}\delta({\bf p}+ {\bf k}),\cr
&&\langle \Phi_0|a^\dagger_\lambda({\bf p},t)\sgn\lambda\,b_{\lambda'}({\bf k})|\Phi_0\rangle =\frac{1}{2}\left(\beta_\tp^+\beta_\tk ^+e^{-i(\delta_ \tp  -\delta_\tk )}e^{-i\Omega_\tp^+t}-\beta_\tp^-\beta_\tk ^-e^{-i(\delta_ \tp  -\delta_\tk )}e^{-i\Omega_\tp^-t}\right)\delta_{\lambda\lambda'}\delta({\bf p}- {\bf k}),\cr
&&\langle \Phi_0|\sgn\lambda\,b_\lambda({\bf p},t)a^\dagger_{\lambda'}({\bf k})|\Phi_0\rangle =\frac{1}{2}\left(\alpha_\tp^+\alpha_\tk ^+e^{-i\Omega_\tp^+t}-\alpha_\tp^-\alpha_\tk ^-e^{-i\Omega_\tp^-t}\right)\delta_{\lambda\lambda'}\delta({\bf p}- {\bf k}).
\end{eqnarray}
Hence, we find the vacuum-to-vacuum transition amplitude which reads in general
\begin{eqnarray}\label{nonrel-prop_gen}
&&\langle \Phi_0|\Psi^c({\bf x},t)\bar\Psi({\bf y},0)|\Phi_0\rangle=\int\frac{d^3p}{(2\pi)^3 2\omega_\tp}\sum_{\lambda}\cr
&&\times\Big[\sgn\lambda\frac{i}{2}\left(\alpha_\tp^+\beta_\tp^+e^{-i\delta_ \tp  }e^{-i\Omega_\tp^+t}+\alpha_\tp^-\beta_\tp^-e^{-i\delta_ \tp  }e^{-i\Omega_\tp^-t}\right)v_\lambda({\bf p})\bar u_{\lambda}(-{\bf p})e^{-i{\bf p}\cdot ({\bf x }-{\bf  y })}\cr
&&-\sgn\lambda\frac{i}{2}\left(\alpha_\tp^+\beta_\tp^+e^{i\delta_ \tp  }e^{-i\Omega_\tp^+t}+\alpha_\tp^-\beta_\tp^-e^{i\delta_ \tp  }e^{-i\Omega_\tp^-t}\right)u_\lambda({\bf p})\bar v_{\lambda}(-{\bf p})e^{i{\bf p}\cdot ({\bf x }-{\bf  y })}\cr
&&+\frac{1}{2}\left((\beta_\tp^+)^2e^{-i\Omega_\tp^+t}-(\beta_\tp^-)^2e^{-i\Omega_\tp^-t}\right)v_\lambda({\bf p})\bar v_{\lambda}({\bf p})e^{-i{\bf p}\cdot ({\bf x }-{\bf  y })}\cr
&&+\frac{1}{2}\left((\alpha_\tp^+)^2e^{-i\Omega_\tp^+t}-(\alpha_\tp^-)^2e^{-i\Omega_\tp^-t}\right)u_\lambda({\bf p})\bar u_{\lambda}({\bf p})e^{i{\bf p}\cdot ({\bf x }-{\bf  y })}
\Big],
\end{eqnarray}
with the coefficients given by \eqref{sc_delta}, \eqref{BT_coeff+} and \eqref{BT_coeff-}.

%%%%%%%%%%%%%%%%%%%%%%%%%%%%%%%%%%%%%%%%%

\subsection{Mass-degenerated Majorana quasiparticles  ($\epsilon_1=0,\epsilon_5\neq0$)}\label{propagator_RBT}

The comparison is more transparent if we consider specific cases. The most interesting and illuminating is the case when $\epsilon_1=0$. Then, the Majorana quasiparticles are degenerate in mass and the neutron-antineutron oscillation does not happen (see Sect. \ref{trans_prob}). However, as discussed in \cite{FT1} and \cite{FT3}, the anomalous propagator derived in the Lagrangian framework is nonvanishing. 

The Lagrangian in this case is:
\begin{eqnarray}\label{Lagr_NO}
{\cal L}&=&\overline{\Psi}(x)i\gamma^{\mu}\partial_{\mu}\Psi(x) - m\overline{\Psi}(x)\Psi(x)\nonumber\\
&-&\frac{i}{2}\epsilon_{5}[\Psi^{T}(x)C\gamma_{5}\Psi(x)  + \overline{\Psi}(x)C\gamma_{5}\overline{\Psi}^{T}(x)],
\end{eqnarray}
leading to the equations of motion:
\begin{eqnarray}\label{47'}
\left(i\gamma^{\mu}\partial_{\mu}-m- i\epsilon_5 \gamma_{5}\right)\Psi_+(x)=0,\cr
\left(i\gamma^{\mu}\partial_{\mu}-m+ i\epsilon_5 \gamma_{5}\right)\Psi_-(x)=0.
\end{eqnarray}
We define
\begin{eqnarray}\label{48}
m\pm i\epsilon_5 \gamma_{5}=Me^{\pm 2i\Theta\gamma_{5}}
\end{eqnarray}
with
\begin{eqnarray}\label{49}
M=\sqrt{m^{2}+\epsilon_5^{2}}.
\end{eqnarray}
The equations of motion can be recast in the form
\begin{eqnarray}\label{50}
\left(i\gamma^{\mu}\partial_{\mu}-M\right)e^{\pm i\Theta\gamma_{5}}(\Psi(x)\mp \Psi^c(x))=0
\end{eqnarray}

We thus identify the combinations of Majorana type
\begin{eqnarray}\label{51}
\tilde\Psi_{+}&=&\frac{1}{\sqrt 2}e^{i\Theta\gamma_{5}}(\Psi(x)- \Psi^c(x)),\nonumber\\
\tilde\Psi_{-}&=&\frac{1}{\sqrt 2}e^{- i\Theta\gamma_{5}}(\Psi(x)+ \Psi^c(x)),
\end{eqnarray}
which satisfy the standard Dirac equation
\begin{eqnarray}\label{52}
\left(i\gamma^{\mu}\partial_{\mu}-M\right)\tilde\Psi_{\pm}(x)=0.
\end{eqnarray}
Thus we have the exact solutions of the field equations,
\begin{eqnarray}\label{55}
&&\Psi(x)=\frac{1}{\sqrt 2}[e^{-i\Theta\gamma_{5}}\tilde\Psi_{+}(x)+e^{i\Theta\gamma_{5}}\tilde\Psi_{-}(x)],\nonumber\\
&&\Psi^c(x)=\frac{1}{\sqrt 2}[e^{-i\Theta\gamma_{5}}\tilde\Psi_{+}(x)-e^{i\Theta\gamma_{5}}\tilde\Psi_{-}(x)].
\end{eqnarray}
The Majorana fields $\tilde\Psi_\pm(x)$ can be also mixed into a Dirac-type of field $N(x)$, with a shifted mass $M=\sqrt{m^{2}+\epsilon^{2}}$:
\begin{eqnarray}\label{56}
N(x)=\frac{1}{\sqrt2} (\tilde\Psi_{+}(x)+ \tilde\Psi_{-}(x)),\cr
N^c(x)= \frac{1}{\sqrt2} (\tilde\Psi_{+}(x)- \tilde\Psi_{-}(x)),
\end{eqnarray}
satisfying in its turn the simple Dirac equation
\begin{eqnarray}\label{Dirac_N}
\left(i\gamma^{\mu}\partial_{\mu}-M\right)N(x)=0.
\end{eqnarray}

We can than rewrite \eqref{55} as 
\begin{eqnarray}\label{RBT}
\left(\begin{array}{c}
            \Psi(x)\\
            \Psi^{c}(x)
            \end{array}\right)
&=& \left(\begin{array}{c}
            \cos\Theta N(x)-i\gamma_{5}\sin\Theta N^c(x)\\
            \cos\Theta N^c(x)-i\gamma_{5}\sin\Theta N(x)
            \end{array}\right).
\end{eqnarray}

This transformation, mixing the relativistic neutron field and its charge conjugated, has been named "relativistic Bogoliubov transformation" \cite{ft, FT1}. It is easy to see that it preserves the anticommutators, therefore it is canonical. In the form quoted above, it is covariant under charge conjugation transformation. It has been used in \cite{FT1} to analyse the neutron oscillations in the Lagrangian description; a charge-conjugation violating version of it has been used in \cite{FT2} to provide a two-step solution to the seesaw mechanism. 

Due to the simplicity of the equation of motion satisfied by the field $N(x)$, the relativistic Bogoliubov transformation \eqref{RBT} leads us immediately to the form of the anomalous propagator for $\epsilon_1=0$:
\begin{equation}\label{anom_prop_1}
\langle T^*\Psi^c(x)\bar\Psi(y)\rangle= \int\frac{d^{4}p}{(2\pi)^{4}}\frac{\epsilon_5\gamma_{5}}{p^{2}-M^{2}+i\epsilon}e^{-ip(x-y)},
\end{equation}
for which we used
\begin{equation}\label{usual_prop}
\langle T^*\Psi(x)\bar\Psi(y)\rangle=\langle T^*\Psi^c(x)\bar{\Psi^c}(y)\rangle =\int\frac{d^{4}p}{(2\pi)^{4}}\frac{i}{\pslash-M+i\epsilon}e^{-ip(x-y)}.
\end{equation}
Recall that the propagator theory based on equation of motion, or the path integral quantization, give the above covariant $T^*$-product, while the canonical quantization leads to the usual $T$-product \eqref{anom_prop}, which specifies precisely the equal-time limit of the correlation. The two products coincide if the $T^*$-product vanishes in the limit $p_0\to\infty$, which is the case also in our situations.

On the one hand, starting with the covariant propagator \eqref{anom_prop_1}, we easily find
\begin{eqnarray}\label{prop_NO_Lagr}
\langle \Phi_0|\Psi^c({\bf x},t)\bar\Psi({\bf y},0)|\Phi_0\rangle&=&\epsilon_5\gamma_5\int\frac{d^3p}{(2\pi)^3}\oint_{\Gamma(\Omega_\tp)}\frac{dp_0}{2\pi}\frac{e^{-ip(x-y)}}{p_0^2-\Omega_\tp^2}\cr
&=&i\epsilon_5\gamma_5\int\frac{d^3p}{(2\pi)^32\Omega_\tp}e^{-i\Omega_ \tp t+i{\bf p}\cdot({\bf x}-{\bf y})},
\end{eqnarray}
where $\Gamma(\Omega_\tp)$ is the contour in the complex $p_0$-plane which includes the pole of the integrand at $\Omega_\tp$ and extends to $-i\infty$ in the lower half-plane.

On the other hand, starting from the general formula \eqref{nonrel-prop_gen} in the Hamiltonian approach, we obtain
\begin{eqnarray}\label{prop_NO_Ham}
\langle \Phi_0|\Psi^c({\bf x},t)\bar\Psi({\bf y},0)|\Phi_0\rangle&=&-\int\frac{d^3p}{(2\pi)^32\omega_\tp}\sum_\lambda\alpha_\tp\beta_\tp[v_\lambda(-{\bf p})\bar u_{\lambda}({\bf p})+u_\lambda({\bf p})\bar v_{\lambda}(-{\bf p})]e^{-i\Omega_ \tp t+i{\bf p}\cdot({\bf x}-{\bf y})}\cr
&=&i\epsilon_5\gamma_5\int\frac{d^3p}{(2\pi)^32\Omega_\tp}e^{-i\Omega_ \tp t+i{\bf p}\cdot({\bf x}-{\bf y})},
\end{eqnarray}
where we specified the coefficients for the case $\epsilon_1=0$ as follows:
\begin{eqnarray}
&&\Omega_\tp^\pm=\Omega_\tp=\sqrt{\tp^2+m^2+\epsilon_5^2},\cr
&&\alpha_\tp^\pm=\alpha_\tp=\sqrt{\frac{\Omega_\tp+\omega_\tp}{2\Omega_\tp}},\cr
&&\beta_\tp^\pm=\beta_\tp=-\sqrt{\frac{\Omega_\tp-\omega_\tp}{2\Omega_\tp}},\cr
&&\sin\delta_ \tp =-\sgn\lambda,\ \ \ \cos\delta_ \tp =0,
\end{eqnarray}
and we also used \eqref{mr1}--\eqref{mr2}.

The coincidence of the results \eqref{prop_NO_Lagr} and \eqref{prop_NO_Ham} indicates the agreement of the Lagrangian approach, with the use of the relativistic Bogoliubov transformation \eqref{RBT}, and the Hamiltonian formalism via Bogoliubov quasiparticles and vacuum condensate developed above. In this case,  the neutron-antineutron conversion takes place in virtual states and it is strictly an effect of the vacuum
condensate, which violates the baryon number conservation. Interestingly, the relativistic Bogoliubov transformation does not have any r\^ole in the Hamiltonian description.

\subsection{Mass non-degenerate Majorana quasiparticles ($\epsilon_1\neq0,\epsilon_5=0$)}

For completeness, we include also the calculation of the anomalous propagator for the typical neutron-antineutron oscillation setup, with $\epsilon_5=0$. In this case, the equations of motion for the primary Majorana fields are simply
\begin{eqnarray}\label{eom_osc}
(i\gamma^{\mu}\partial_{\mu}-M_+)\Psi_+(x)=0,\cr
(i\gamma^{\mu}\partial_{\mu}-M_-)\Psi_-(x)=0,
\end{eqnarray}
with $M_\pm=m\pm\epsilon_1$.
The anomalous propagator of the neutron field is easily obtained with the help of the relations \eqref{Majorana} and the ordinary propagators
\begin{equation}\label{ord_prop}
\langle T^*\Psi_\pm(x)\bar\Psi_\pm(y)\rangle =\int\frac{d^{4}p}{(2\pi)^{4}}\frac{i}{\pslash-M_\pm+i\epsilon}e^{-ip(x-y)},
\end{equation}
as
\begin{equation}\label{ord_prop}
\langle T^*\Psi^c(x)\bar\Psi(y)\rangle =\frac{1}{2}\int\frac{d^{4}p}{(2\pi)^{4}}\left(\frac{1}{\pslash-M_++i\epsilon}-\frac{1}{\pslash-M_-+i\epsilon} \right)e^{-ip(x-y)}.
\end{equation}
We note that %
\begin{eqnarray}\label{prop_osc_Lagr}
\langle \Phi_0|\Psi_\pm({\bf x},t)\bar\Psi_\pm({\bf y},0)|\Phi_0\rangle&=&i\int\frac{d^3p}{(2\pi)^3}\oint_{\Gamma(\Omega_\tp^\pm)}\frac{dp_0}{2\pi}e^{-ip(x-y)}\frac{\pslash+M_\pm}{p_0^2-(\Omega_\tp^\pm)^2}\\
&=&-\int\frac{d^3p}{(2\pi)^32\Omega_\tp^\pm}e^{-i\Omega_\tp^\pm t+i{\bf p}\cdot({\bf x}-{\bf y})}\left(\Omega_\tp^\pm\gamma_0-{\bf p}\cdot{\bf \gamma}+M_\pm\right).\nonumber
\end{eqnarray}
We shall consider the above amplitudes in the limit $\epsilon_1\ll m$ and in the first order in $\epsilon_1$. In this approximation,
$$
\frac{1}{2\Omega_\tp^\pm}\left(\Omega_\tp^\pm\gamma_0-{\bf p}\cdot{\bf \gamma}+M_\pm\right)=\frac{1}{2\omega_\tp}\left[(\omega_\tp\gamma_0-{\bf p}\cdot{\bf \gamma}+m)\pm\epsilon_1\left(\frac{\tp^2}{\omega_\tp^2 }+\frac{m}{\omega_\tp^2}{\bf p}\cdot{\bf \gamma}\right)\right].
$$
Then,
\begin{eqnarray}\label{prop_osc_Lagr_1}
\langle \Phi_0|\Psi^c({\bf x},t)\bar\Psi({\bf y},0)|\Phi_0\rangle=\int\frac{d^3p}{(2\pi)^32\omega_\tp}e^{-i\omega_\tp t+i{\bf p}\cdot({\bf x}-{\bf y})}\Big[i\sin\left(\frac{\epsilon_1m}{\omega_\tp}t\right)(\omega_\tp\gamma_0-{\bf p}\cdot{\bf \gamma}+m)\cr
-\epsilon_1\cos\left(\frac{\epsilon_1 m}{\omega_\tp}t\right)\left(\frac{\tp^2}{\omega_\tp^2 }+\frac{m}{\omega_\tp^2}{\bf p}\cdot{\bf \gamma}\right)\Big].
\end{eqnarray}

We calculate now the same amplitude starting from the canonical quantization result \eqref{nonrel-prop_gen}, up to the first order in $\epsilon_1$, in which case
\begin{eqnarray}
\left(\alpha_\tp^\pm\right)^2\approx 1-\epsilon_1^2\frac{\tp^2}{4\omega_\tp^4},\ \ \ \ 
\left(\beta_\tp^\pm\right)^2\approx\epsilon_1^2\frac{\tp^2}{4\omega_\tp^4},\ \ \ \ 
\alpha_\tp^\pm\beta_\tp^\pm\approx \epsilon_1\frac{\tp}{2\omega_\tp^2}
\end{eqnarray}
and
\begin{eqnarray}
\sin\delta_ \tp =0,\ \ \ \cos\delta_ \tp   =1.
\end{eqnarray}
We obtain, using \eqref{mr3}, \eqref{mr4}:
\begin{eqnarray}\label{nonrel-prop_osc}
\langle \Phi_0|\Psi^c({\bf x},t)\bar\Psi({\bf y},0)|\Phi_0\rangle&=&i\int\frac{d^3p}{(2\pi)^3 2\omega_\tp}e^{-i\omega_\tp t+i{\bf p}\cdot ({\bf x }-{\bf  y })}\cr
&\times&\sum_{\lambda}\Big[\epsilon_1\frac{\tp}{2\omega_\tp^2}\cos\left(\frac{\epsilon_1m}{\omega_\tp}t\right)\sgn\lambda\left( v_\lambda({\bf p})\bar u_{\lambda}(-{\bf p})
-u_\lambda({\bf p})\bar v_{\lambda}(-{\bf p})\right)\cr
&-&\sin\left(\frac{\epsilon_1m}{\omega_\tp}t\right)u_\lambda({\bf p})\bar u_{\lambda}({\bf p})\Big]\cr
&=&\int\frac{d^3p}{(2\pi)^32\omega_\tp}e^{-i\omega_\tp t+i{\bf p}\cdot({\bf x}-{\bf y})}\Big[i\sin\left(\frac{\epsilon_1m}{\omega_\tp}t\right)(\omega_\tp\gamma_0-{\bf p}\cdot{\bf \gamma}+m)\cr
&-&\epsilon_1\cos\left(\frac{\epsilon_1 m}{\omega_\tp}t\right)\left(\frac{\tp^2}{\omega_\tp^2 }+\frac{m}{\omega_\tp^2}{\bf p}\cdot{\bf \gamma}\right)\Big].
\end{eqnarray}
This coincides with the formula \eqref{prop_osc_Lagr_1}, and thus the two approaches prove again compatible.

%%%%%%%%%%%%%%%%%%%%%%%%%%%%%%%%%%%%

\section{Discussion and conclusions}

The phenomena of particle oscillations are highly peculiar, in the sense that the particles which are supposed to oscillate do not exist as well-defined states in the quantum field theory. It is perhaps more accurate to speak about the oscillations of a flavour quantum number than about the oscillations of particles. Actually, one of the recurring questions in the theory of neutrino oscillations is how to define the {\it flavour neutrino states}, when the flavour fields are given as known mixtures of massive fields. The answer can be found by using the canonical quantization procedure described in this work. Specifically, in the baryon-number violating case of the neutron-antineutron oscillations, the neutron field $\Psi(x)$ is a mixing of massive Majorana fields $\Psi_\pm(x)$ (which diagonalize the Lagrangian \eqref{Lagr_osc}) according to eq. \eqref{neutron_field_mixing}:
$$\Psi(x)=\frac{1}{\sqrt2}\big(\Psi_+(x)+ \Psi_-(x)\big).$$
On the other hand, we have found that the neutron and antineutron states associated with  $\Psi (x)$ are mixings of states associated with  $\Psi_\pm(x)$ according to eqs. \eqref{n_state} and \eqref{nbar_state}:
\begin{eqnarray*}
|n({\bf p},\lambda)\rangle&=&
\frac{1}{\sqrt 2}\left[\left(\frac{\Omega_\tp^++\omega_\tp}{2\Omega_\tp^+}+\frac{m\epsilon_1}{2\omega_\tp\Omega_\tp^+}\right)A^\dagger_\lambda({\bf p})+\left(\frac{\Omega_\tp^-+\omega_\tp}{2\Omega_\tp^-}-\frac{m\epsilon_1}{2\omega_\tp\Omega_\tp^-}\right)B^\dagger_\lambda({\bf p})\right]|\Phi_0\rangle,\\
|\bar n({\bf p},\lambda)\rangle&=&\frac{1}{\sqrt 2}\sgn\,\lambda \left[\left(\frac{\Omega_\tp^++\omega_\tp}{2\Omega_\tp^+}+\frac{m\epsilon_1}{2\omega_\tp\Omega_\tp^+}\right)A^\dagger_\lambda({\bf p})-\left(\frac{\Omega_\tp^-+\omega_\tp}{2\Omega_\tp^-}-\frac{m\epsilon_1}{2\omega_\tp\Omega_\tp^-}\right)B^\dagger_\lambda({\bf p})\right]|\Phi_0\rangle,
\end{eqnarray*}
where $|\Phi_0\rangle$ is the physical vacuum of the model, $A^\dagger$ and $B^\dagger$ are the creation operators corresponding to the Majorana fields $\Psi_+$ and $\Psi_-$, respectively, and $\omega_\tp=\sqrt{\tp^2+m^2}$,  $\Omega_\tp^\pm=\sqrt{\tp^2+(m\pm\epsilon_1)^2+\epsilon_5^2}$.
This mixing of states is not self-evident just by knowing the mixing of fields, but it is unambiguous, once the appropriate quantization scheme is employed.

The present canonical quantization approach to the neutron-antineutron oscillations is based on the theory of unitarily inequivalent representations inspired by the BCS theory of superconductivity \cite{BCS, Bogoliubov} and the Nambu--Jona-Lasinio model \cite{NJL} (see also \cite{UTK}). In this formulation, the physical particles are Bogoliubov quasiparticles with definite masses. The neutron field is a mixing of two mass-nondegenerate Majorana fields, consequently it cannot be a proper Dirac field. Flavour (in this case, baryonic number) is introduced into the theory by the {\it would-be neutron fields} in the absence of the baryon-number violating interaction. These fields provide one of the two inequivalent representations of the creation and annihilation operators used in the analysis, and it should be emphasized that it is a nonphysical representation.

The intuitive picture is the following: the physical free Majorana particles, or Bogoliubov quasiparticles, are states in a Fock space, built on a vacuum which is a coherent superposition of bosonic pairs of bare neutrons and antineutrons, with opposite momenta and spins. The vacuum violates baryon number conservation, since the pairs carry  baryonic number $\pm2$. In this Fock space, the neutron and antineutron are defined using only the physical vacuum and the neutron field in the Lagrangian, which gives the dynamical relation of the neutron to the other particles in the Standard Model. Neutron and antineutron "states" are superpositions of Bogoliubov quasiparticles, satisfying the consistency requirement that, in the limit of vanishing baryon-number violating interactions, they are identical to the Standard Model neutron and antineutron states. The oscillation takes place due to the mass splitting between the Majorana particles, which is an effect of the vacuum condensate, that acts also as a reservoir of baryonic number.

The analogy with the BCS theory is only partial. The BCS ground state is a state of matter -- the interaction couples the electrons in Cooper pairs and provides the {\it energy gap} which renders the system superconducting. The physical particles are the same, i.e. electrons with a well-defined mass, with or without the interaction.  On the other hand, in relativistic models, the only physical states are quasiparticle states. The bare (or flavour) particles simply do not exist, either individually or in pairs. This can be seen easily by the fact that we could have chosen initially a different unitarily inequivalent representation than the one corresponding to the bare particles and the result of diagonalization would have been exactly the same. Hence, the vacuum condensate is a technical device which mimics the effect of the interaction in breaking the baryon-number symmetry and generating the mass splitting and/or the mass gap for the quasiparticles.

One of the lessons learned from this analysis is that the identification of the primary fields is influenced by the conserved symmetries of the model. For instance, in the case of neutron-antineutron oscillations considered here, the Lagrangian \eqref{Lagr_osc} is C-invariant, and the primary fields are C-eigenfields. In the case of the seesaw mechanism for neutrinos, when charge conjugation is violated but the lepton number violation is still given by Majorana mass terms, the C-violation has to be absorbed by the vacuum, in order for the physical fields to be eigenfields of charge conjugation. This is achieved in the Lagrangian approach by the relativistic Bogoliubov transformation \cite{FT, FT2}. Recently, an alternative analysis of the vacuum condensate, based on the relativistic Bogoliubov transformation, was performed in \cite{KF} for the C-violating model of the seesaw mechanism for neutrinos (whose Lagrangian is \eqref{1}, with $\alpha=\pi/2$). It would be interesting to apply the formulation of the present work also for the seesaw mechanism \cite{work_in_progress}. 

The quantization by path integral methods in the Lagrangian formalism \cite{FT3} and the present canonical formalism converge, as it was shown by comparing the anomalous propagators obtained in the two approaches. In our canonical description, the relativistic Bogoliubov transformation which is used for the (partial) diagonalization of the Lagrangian \cite{FT1, FT3} does not play any role. 
The advantage of the canonical formalism is that it allows us to define the neutron and antineutron "states" and the vacuum condensate offers a richer picture. Flavour states are defined only as auxiliary notions, which bring the baryonic number into the picture, and their Fock space is unphysical. An alternative procedure in the context of neutrino mixing has been developed in \cite{BV1} (see also \cite{BV2} and references therein, and \cite{CG} as well for a critique of the method), invoking unitarily inequivalent representations, but in which the flavour Fock space features as a physical space. The technical details are at variance with the approach presented in our work.

The method of quantization for oscillating particle systems described in this work can be easily applied to mixings of Dirac neutrinos. Regarding the seesaw mechanism, a possible extension of the method will provide also a clarification of the charge conjugation violation by the vacuum condensate. We plan to study these issues elsewhere.

\subsection*{Acknowledgments}
I am grateful to Kazuo Fujikawa for many useful discussions. I would like to thank Masud Chaichian and Archil Kobakhidze for several clarifying comments.

\appendix

\section{Conventions for spinors}\label{appendix1}

We work with the Dirac representation of the $\gamma$-matrices:
\begin{eqnarray}\label{gamma_matr}
\gamma^0=\left(\begin{array}{cc}
            \sigma^0&0\\
            0&-\sigma^0
            \end{array}\right),\ \ \ \ 
\gamma^i=\left(\begin{array}{cc}
            0&\sigma^i\\
            \sigma^i&0
            \end{array}\right),\ \ \ \
\gamma_5=\left(\begin{array}{cc}
            0&\sigma^0\\
            \sigma^0&0
            \end{array}\right),
\end{eqnarray}
where $\sigma^0={\mathbf 1}_{2\times 2}$ and $\sigma^i$, $i=1,2,3$ are the Pauli matrices.

The solution of the Dirac equation
\begin{eqnarray}\label{D_eq}
(i\gamma^\mu\partial_\mu-m)\psi(x)=0
\end{eqnarray}
is written in mode expansion as
\begin{eqnarray}\label{mode_exp}
\psi(x)=\int\frac{d^3p}{(2\pi)^{3/2}\sqrt{2\omega_\tp}}\sum_\lambda\left(a_\lambda({\bf p})u_\lambda({\bf p})e^{-ipx}+b^\dagger_\lambda({\bf p})v_\lambda({\bf p})e^{ipx}\right),
\end{eqnarray}
where $\lambda=\pm{\frac{1}{2}}$ are the helicity eigenvalues and $p_0=\omega_\tp=\sqrt{\tp^2+m^2}$, with the notation $\text{p}=|{\bf p}|$. The spinors $u_\lambda({\bf p})$ and $v_\lambda({\bf p})$ are helicity eigenvectors,
\begin{eqnarray}
\frac{\hat{\bf S}\cdot{\bf p}}{\tp}u_\lambda({\bf p})=\lambda u_\lambda({\bf p}),\ \ \ \frac{\hat{\bf S}\cdot{\bf p}}{\tp}v_\lambda({\bf p})=-\lambda v_\lambda({\bf p}),
\end{eqnarray}
with the spin matrix
\begin{eqnarray}
\hat{\bf S}^i= \frac{1}{2}{\bf \Sigma}^i=\left(\begin{array}{cc}
            {\sigma}^i&0\\
            0&\sigma^i
            \end{array}\right).
\end{eqnarray}

The left- and right-handed helicity spinors read:
\begin{eqnarray}\label{helicity_spinors}
u_{\uparrow}({\bf p})=\sqrt{\omega_\tp+m}\left(\begin{array}{c}
            \chi_\uparrow\\
            \frac{\text{p}}{\omega_\tp+m}\chi_\uparrow
            \end{array}\right)\ ,\ \ \ 
u_{\downarrow}({\bf p})=\sqrt{\omega_\tp+m}\left(\begin{array}{c}
            \chi_\downarrow\\
            -\frac{\text{p}}{\omega_\tp+m}\chi_\downarrow
            \end{array}\right)\ ,\cr
v_{\uparrow}({\bf p})=\sqrt{\omega_\tp+m}\left(\begin{array}{c}
           -\frac{\text{p}}{\omega_\tp+m}\eta_\uparrow\\
            \eta_\uparrow
            \end{array}\right)\ ,\ \ \ 
v_{\downarrow}({\bf p})=\sqrt{\omega_\tp+m}\left(\begin{array}{c}
            \frac{\text{p}}{\omega_\tp+m}\eta_\downarrow\\
            \eta_\downarrow
            \end{array}\right)\ ,
\end{eqnarray}
where the symbol $\uparrow$ denotes the right-handed spinor, while $\downarrow$ denotes the left-handed spinor.
We use the helicity basis 
\begin{eqnarray}\label{helicity_basis}
\chi_{\uparrow}=\eta_\downarrow=\left(\begin{array}{c}
            \cos\frac{\theta}{2}e^{-i\frac{\phi}{2}}\\
             \sin\frac{\theta}{2}e^{i\frac{\phi}{2}}
            \end{array}\right)\ ,\ \ \ 
\chi_{\downarrow}=\eta_\uparrow=\left(\begin{array}{c}
            -\sin\frac{\theta}{2}e^{-i\frac{\phi}{2}}\\
             \cos\frac{\theta}{2}e^{i\frac{\phi}{2}}
            \end{array}\right),
\end{eqnarray}
with $\theta$ and $\phi$ being the polar and azimuthal angles of the momentum vector, \\${\bf p}=(\text{p}\sin\theta\cos\phi,\ \text{p}\sin\theta\sin\phi,\ \text{p}\cos\theta)$. The basis spinors $\chi_\lambda$ and $\eta_\lambda$ satisfy
\begin{eqnarray}
({\vec \sigma}\cdot{\bf p})\chi_\lambda=2\lambda\,\tp\, \chi_\lambda,  \ \ \ \ ({\vec \sigma}\cdot{\bf p})\eta_\lambda=-2\lambda\,\tp\,\eta_\lambda
\end{eqnarray}
and are normalized as
\begin{eqnarray}\label{basis}
\chi^\dagger_\lambda\chi_{\lambda'}=\eta^\dagger_\lambda\eta_{\lambda'}=\delta_{\lambda\lambda'}, \ \rm{where}\ \  \lambda, \lambda'=\pm\frac{1}{2}.
\end{eqnarray}
We note as well the relations:
\begin{eqnarray}
&&\chi_\uparrow\chi^\dagger_\uparrow+\chi_\downarrow\chi^\dagger_\downarrow={\mathbf 1}_{2\times2},\\
&&\chi_\uparrow\chi^\dagger_\uparrow-\chi_\downarrow\chi^\dagger_\downarrow=\left(\begin{array}{cc}
            \cos{\theta}&\sin\theta e^{-i{\phi}}\\
             \sin{\theta}e^{i{\phi}}&-\cos\theta
            \end{array}\right)=\frac{{\vec \sigma}\cdot{\bf p}}{\tp}.
\end{eqnarray}

The helicity spinors for the inverted momentum vector $-\bf p$ are obtained by taking $\theta\to\pi-\theta$ and $\phi\to\pi+\phi$ in \eqref{helicity_spinors} and they read as follows:
\begin{eqnarray}\label{helicity_spinors_minus}
u_{\uparrow}(-{\bf p})=i\sqrt{\omega_\tp+m}\left(\begin{array}{c}
            \eta_\uparrow\\
            \frac{\text{p}}{\omega_\tp+m}\eta_\uparrow
            \end{array}\right)\ ,\ \ \ 
u_{\downarrow}(-{\bf p})=i\sqrt{\omega_\tp+m}\left(\begin{array}{c}
            \eta_\downarrow\\
            -\frac{\text{p}}{\omega_\tp+m}\eta_\downarrow
            \end{array}\right)\ ,\cr
v_{\uparrow}(-{\bf p})=i\sqrt{\omega_\tp+m}\left(\begin{array}{c}
           -\frac{\text{p}}{\omega_\tp+m}\chi_\uparrow\\
            \chi_\uparrow
            \end{array}\right)\ ,\ \ \ 
v_{\downarrow}(-{\bf p})=i\sqrt{\omega_\tp+m}\left(\begin{array}{c}
            \frac{\text{p}}{\omega_\tp+m}\chi_\downarrow\\
            \chi_\downarrow
            \end{array}\right)\ .
\end{eqnarray}

The helicity spinors  are normalized as
\begin{eqnarray}\label{spinor_norm}
u^\dagger_\lambda({\bf p})u_{\lambda'}({\bf p})&=&2\omega_\tp\delta_{\lambda\lambda'},\cr
u^\dagger_\lambda({\bf p})v_{\lambda'}(-{\bf p})&=&0,
\end{eqnarray}
and
satisfy the relations:
\begin{eqnarray}\label{spinor_rel}
\bar u_\lambda({\bf p})u_{\lambda'}({\bf p})&=&2m\delta_{\lambda\lambda'},\cr
\bar v_\lambda({\bf p})v_{\lambda'}({\bf p})&=&-2m\delta_{\lambda\lambda'},\cr
\bar u_\lambda({\bf p})v_{\lambda'}(-{\bf p})&=&-2i{\text p}\,{\text {sgn}}\lambda\, \delta_{\lambda\lambda'},\cr
\bar v_\lambda({\bf p})u_{\lambda'}(-{\bf p})&=&-2i{\text p}\,{\text {sgn}}\lambda\, \delta_{\lambda\lambda'},
\end{eqnarray}
as well as
\begin{eqnarray}\label{spinor_rel_5}
\bar u_\lambda({\bf p})\gamma_5 u_{\lambda'}({\bf p})&=&0,\cr
\bar v_\lambda({\bf p})\gamma_5v_{\lambda'}({\bf p})&=&0,\cr
\bar u_\lambda({\bf p})\gamma_5 v_{\lambda'}(-{\bf p})&=&2i\omega_\tp \delta_{\lambda\lambda'},\cr
\bar v_\lambda({\bf p})\gamma_5 u_{\lambda'}(-{\bf p})&=&-2i\omega_\tp \delta_{\lambda\lambda'}.
\end{eqnarray}
We also have the relations:
\begin{eqnarray}
\sum_{\lambda}v_\lambda(-{\bf p})\bar u_{\lambda}({\bf p})&=&i(\omega_\tp+m)\left(\begin{array}{cc}
            -\frac{\tp}{\omega_\tp+m}\frac{{\vec \sigma}\cdot{\bf p}}{\tp}&\frac{\tp^2}{(\omega_\tp+m)^2}{\mathbf 1}_{2\times2}\\
             {\mathbf 1}_{2\times2}&-\frac{\tp}{\omega_\tp+m}\frac{{\vec \sigma}\cdot{\bf p}}{\tp}
            \end{array}\right),\label{mr1}\\
\sum_{\lambda}u_\lambda({\bf p})\bar v_{\lambda}(-{\bf p})&=&i(\omega_\tp+m)\left(\begin{array}{cc}
            \frac{\tp}{\omega_\tp+m}\frac{{\vec \sigma}\cdot{\bf p}}{\tp}&{\mathbf 1}_{2\times2}\\
          \frac{\tp^2}{(\omega_\tp+m)^2}  {\mathbf 1}_{2\times2}&\frac{\tp}{\omega_\tp+m}\frac{{\vec \sigma}\cdot{\bf p}}{\tp}
            \end{array}\right),\label{mr2}\\
\sum_{\lambda}\sgn \lambda\,v_\lambda(-{\bf p})\bar u_{\lambda}({\bf p})&=&i(\omega_\tp+m)\left(\begin{array}{cc}
            -\frac{\tp}{\omega_\tp+m}{\mathbf 1}_{2\times2}&\frac{\tp^2}{(\omega_\tp+m)^2}\frac{{\vec \sigma}\cdot{\bf p}}{\tp}\\
             \frac{{\vec \sigma}\cdot{\bf p}}{\tp}& -\frac{\tp}{\omega_\tp+m}{\mathbf 1}_{2\times2}
            \end{array}\right),\label{mr3}\\
\sum_{\lambda}\sgn \lambda\,u_\lambda({\bf p})\bar v_{\lambda}(-{\bf p})&=&i(\omega_\tp+m)\left(\begin{array}{cc}
            \frac{\tp}{\omega_\tp+m}{\mathbf 1}_{2\times2}&\frac{{\vec \sigma}\cdot{\bf p}}{\tp}\\
             \frac{\tp^2}{(\omega_\tp+m)^2}\frac{{\vec \sigma}\cdot{\bf p}}{\tp}& \frac{\tp}{\omega_\tp+m}{\mathbf 1}_{2\times2}
            \end{array}\right).\label{mr4}
\end{eqnarray}

Under the parity transformation,
\begin{eqnarray}\label{spinor_parity}
\gamma_0{u}_\lambda({\bf p})&=&i u_{-\lambda}(-{\bf p}),\cr
\gamma_0{v}_\lambda({\bf p})&=&-i v_{-\lambda}(-{\bf p}),
\end{eqnarray}
such that the parity operation acts as
\begin{eqnarray}\label{psi_parity}
{\cal P}\psi({\bf x},t){\cal P}^{-1}=\gamma_0\psi(-{\bf x},t),
\end{eqnarray}
if 
\begin{eqnarray}\label{annih_parity}
{\cal P}a_\lambda({\bf p},t){\cal P}^{-1}=ia_{-\lambda}(-{\bf p},t),\cr
{\cal P}b_\lambda({\bf p},t){\cal P}^{-1}=ib_{-\lambda}(-{\bf p},t).
\end{eqnarray}

Under the classical charge conjugation transformation, we have 
\begin{eqnarray}\label{spinor_cc}
C\bar{u}_\lambda^T({\bf p})&=&\sgn\lambda\, v_\lambda({\bf p}),\cr
C\bar{v}_\lambda^T({\bf p})&=&\sgn\lambda\, u_\lambda({\bf p}),
\end{eqnarray}
such that 
\begin{eqnarray}\label{charge_conj_4dim}
{\cal C}\psi(x){\cal C}^{-1}=C\bar\psi^T(x)=\int\frac{d^3p}{(2\pi)^{3/2}\sqrt{2\omega_\tp}}\sum_\lambda\text{sgn}\,\lambda\left(b_\lambda({\bf p})u_\lambda({\bf p})e^{-i{p x }}+a^\dagger_\lambda({\bf p})v_\lambda({\bf p})e^{i{p x }}\right),
\end{eqnarray}
with
\begin{eqnarray}
{\cal C}a_\lambda({\bf p}){\cal C}^{-1}= \text{sgn}\,\lambda\, b_\lambda({\bf p}),\ \ \ \ {\cal C}b_\lambda({\bf p}){\cal C}^{-1}= \text{sgn}\,\lambda \, a_\lambda({\bf p}).
\end{eqnarray}

\section{Bogoliubov coefficients from solutions of equations of motion}\label{NJL_calculation}

As explained in Sect. \ref{primary Majoranas}, the Bogoliubov transformations can be found alternatively by equating the explicit solutions of the equations of motion governed by the free Hamiltonian $H_0$ and by the total Hamiltonian $H$. This procedure was used by Nambu and Jona-Lasinio \cite{NJL} (see also \cite{Haag}) for going from massless to massive free Dirac fields. In this Appendix we confirm that this alternative procedure works in the case of the baryon-number violating system with C invariance described by the Lagrangian \eqref{Lagr_osc}, as long as we apply it to primary Majorana fields and not to the mixed field $\Psi(x)$. 

For the sake of transparency, we shall consider two simpler cases, $\epsilon_5=0$ and $\epsilon_1=0$, respectively. In the first case, we use a top-down approach, which consists in imposing the boundary condition and deriving the Bogoliubov transformation. In the second case, we use a bottom-up approach, proving the compatibility of the Bogoliubov transformations (considered to be known) with the boundary condition and the solutions of the equations of motion.

\subsection{Primary fields with scalar Majorana mass ($\epsilon_1\neq0,\epsilon_5=0$)}

The primary fields are Majorana fields, satisfying the free Dirac equations
\begin{eqnarray}\label{eom_osc_1}
(i\gamma^{\mu}\partial_{\mu}-M_+)\Psi_+(x)=0,\cr
(i\gamma^{\mu}\partial_{\mu}-M_-)\Psi_-(x)=0,
\end{eqnarray}
with $M_\pm=m\pm\epsilon_1$.

The natural identification of fields in the Schr\"odinger picture is then
\begin{equation}\label{NJL_osc}
\Psi_\pm({\bf x},0)=\psi_\pm({\bf x},0),
\end{equation}
where $\Psi_\pm({\bf x},0)$ satisfy the  equations of motion \eqref{eom_osc_1}, and $\psi_\pm({\bf x},0)$ satisfy the Dirac equation with mass $m$, having the expressions \eqref{Majorana_mode_exp}. These solutions are easy to find and quantize. For example,
\begin{eqnarray}\label{modes_NJL_osc}
\Psi_+({\bf x},0)&=&\int\frac{d^3p}{(2\pi)^{3/2}\sqrt{2\Omega^+_p}}e^{i{\bf p\cdot x }}\sum_\lambda\left(A_{\lambda}({\bf p})U_\lambda({\bf p})+\sgn\,\lambda\ A^\dagger_{\lambda}(-{\bf p})V_\lambda(-{\bf p})\right)\cr
\psi_+({\bf x},0)&=&\int\frac{d^3p}{(2\pi)^{3/2}\sqrt{2\omega_\tp}}e^{i{\bf p\cdot x }}\sum_\lambda\left(a_{M\lambda}({\bf p})u_\lambda({\bf p})+\sgn\,\lambda\ a^\dagger_{M\lambda}(-{\bf p})v_\lambda(-{\bf p})\right),
\end{eqnarray}
where 
\begin{eqnarray}
(\pslash-M_+)U_\lambda({\bf p})=0,\cr
(\pslash+M_+)V_\lambda({\bf p})=0,
\end{eqnarray}
and we shall consider them of the form \eqref{helicity_spinors}, with $m$ replaced by $M_+$ and $\omega_\tp$ replaced by $\Omega_\tp^+$. Using \eqref{NJL_osc} and \eqref{modes_NJL_osc}, we find
\begin{eqnarray}\label{NJL_osc_eq_coeff}
A_\lambda({\bf p})=\frac{1}{2\sqrt{\Omega_\tp^+\omega_\tp}}U^\dagger_\lambda({\bf p})u_\lambda({\bf p})a_{M\lambda}({\bf p})
+\sgn\lambda\,\frac{1}{2\sqrt{\Omega_\tp^+\omega_\tp}}U^\dagger_\lambda({\bf p})v_\lambda(-{\bf p})a^\dagger_{M\lambda}(-{\bf p}).
\end{eqnarray}
Comparing this expression with \eqref{Ansatz_1}, we infer that \eqref{NJL_osc_eq_coeff} represents the Bogoliubov transformation for $\epsilon_5=0$. Let us calculate the coefficient of $a_{M\lambda}({\bf p})$ in \eqref{NJL_osc_eq_coeff}, using \eqref{helicity_spinors} with the appropriate adjustments for $U_\lambda({\bf p})$:
\begin{eqnarray}
\frac{1}{2\sqrt{\Omega_\tp^+\omega_\tp}}U^\dagger_\lambda({\bf p})u_\lambda({\bf p})&=&\frac{1}{2\sqrt{\Omega_\tp^+\omega_\tp}}\frac{(\Omega_\tp^++M_+)(\omega_\tp+m)+\tp^2}{\sqrt{(\Omega_\tp^++M_+)(\omega_\tp+m)}}\cr
&=&\frac{1}{\sqrt{2\Omega_\tp^+\omega_\tp}}\frac{(\Omega_\tp^++M_++\omega_\tp-m)(\omega_\tp+m)}{\sqrt{2(\Omega_\tp^++M_+)(\omega_\tp+m)}}\cr
&=&\frac{1}{\sqrt{2\Omega_\tp^+\omega_\tp}}\sqrt{\frac{1}{2}\left[(\omega_\tp+m)(\Omega_\tp^++\omega_\tp+\epsilon_1)+(\omega_\tp-m)(\Omega_\tp^++\omega_\tp-\epsilon_1)\right]}\cr
&=&\sqrt{\frac{\Omega_\tp^++\omega_\tp}{2\Omega_\tp^+}+\frac{m\epsilon_1}{2\Omega_\tp^+\omega_\tp}},
\end{eqnarray}
which is indeed identical to $\alpha_\tp^+$ from \eqref{BT_coeff+}, for $\epsilon_5=0$. All the other coefficients will be similarly found to agree with those obtained by the method of Hamiltonian diagonalization in \eqref{BT_coeff+} and \eqref{BT_coeff-}. For $m=0$, one obtains the Bogoliubov coefficients of the Nambu--Jona-Lasinio model \cite{NJL}, with the distinction that in that case the vacuum condensate is formed by pairs of massless particles (nucleons) and antiparticles of opposite spin and momenta, since the usual $U(1)$ symmetry is preserved by the Lagrangian at all stages of the analysis.

\subsection{Primary fields with pseudoscalar Majorana mass ($\epsilon_1=0,\epsilon_5\neq0$)}

Now we consider the primary Majorana fields satisfying the equations of motion
\begin{eqnarray}\label{eom_no_NJL}
[i\gamma^{\mu}\partial_{\mu}-(m+ i\epsilon \gamma_{5})]\Psi_+(x)=0,\cr
[i\gamma^{\mu}\partial_{\mu}-(m- i\epsilon \gamma_{5})]\Psi_-(x)=0.
\end{eqnarray}
We expand $\Psi_+({\bf x},0)$ according to \eqref{Majorana_mode_exp_H} with $\epsilon_1=0$, as 
\begin{eqnarray}
\Psi_+({\bf x},0)&=&\int\frac{d^3p}{(2\pi)^{3/2}\sqrt{2\Omega_\tp}}e^{i{\bf p\cdot x }}\sum_\lambda\left(A_{\lambda}({\bf p})U_\lambda({\bf p})+\sgn\,\lambda\ A^\dagger_{\lambda}(-{\bf p})V_\lambda(-{\bf p})\right),
\end{eqnarray}
where $\Omega_\tp=\sqrt{\tp^2+m^2+\epsilon_5^2}$ and
\begin{eqnarray}\label{Dirac_no_p}
(\pslash-(m+i\epsilon_5\gamma_5))U_\lambda({\bf p})=0,\cr
(\pslash+(m+i\epsilon_5\gamma_5))V_\lambda({\bf p})=0.
\end{eqnarray}
Writing
$$
U_\lambda({\bf p})=\left(\begin{array}{c}
            U_A({\bf p})\\
            U_B({\bf p})
            \end{array}\right),\ \ \ \ 
V_\lambda({\bf p})=\left(\begin{array}{c}
            V_A({\bf p})\\
            V_B({\bf p})
            \end{array}\right)
$$
we find from \eqref{Dirac_no_p} that
\begin{eqnarray}
U_B({\bf p})&=&\frac{{\vec \sigma}\cdot{\bf p}-i\epsilon_5}{\Omega_\tp+m}U_A({\bf p}),\cr
V_A({\bf p})&=&\frac{{\vec \sigma}\cdot{\bf p}-i\epsilon_5}{\Omega_\tp+m}V_B({\bf p}),
\end{eqnarray}
or 
\begin{eqnarray}\label{cond_no_spinor}
U_B({\bf p})=\frac{2\lambda \tp-i\epsilon_5}{\Omega_\tp+m}U_A({\bf p}),\cr
V_A({\bf p})=\frac{-2\lambda\tp-i\epsilon_5}{\Omega_\tp+m}V_B({\bf p}),
\end{eqnarray}
if we require the spinors $U_\lambda({\bf p})$ and $V_\lambda({\bf p})$ to be eigenvectors of the helicity operator as well.

Similarly, $\Psi_-({\bf x},0)$ has the mode expansion
\begin{eqnarray}
\Psi_-({\bf x},0)&=&\int\frac{d^3p}{(2\pi)^{3/2}\sqrt{2\Omega_\tp}}e^{i{\bf p\cdot x }}\sum_\lambda\left(B_{\lambda}({\bf p})\tilde U_\lambda({\bf p})-\sgn\,\lambda\ B^\dagger_{\lambda}(-{\bf p})\tilde V_\lambda(-{\bf p})\right),
\end{eqnarray}
where
\begin{eqnarray}\label{Dirac_no_p-}
(\pslash-(m-i\epsilon_5\gamma_5))\tilde U_\lambda({\bf p})=0,\cr
(\pslash+(m-i\epsilon_5\gamma_5))\tilde V_\lambda({\bf p})=0.
\end{eqnarray}
In this case, for
$$
\tilde U_\lambda({\bf p})=\left(\begin{array}{c}
            \tilde U_A({\bf p})\\
            \tilde U_B({\bf p})
            \end{array}\right),\ \ \ \ 
\tilde V_\lambda({\bf p})=\left(\begin{array}{c}
            \tilde V_A({\bf p})\\
            \tilde V_B({\bf p})
            \end{array}\right)
$$
we find from \eqref{Dirac_no_p-} that
\begin{eqnarray}
\tilde U_B({\bf p})=\frac{{\vec \sigma}\cdot{\bf p}+i\epsilon_5}{\Omega_\tp+m}\tilde U_A({\bf p}),\cr
\tilde V_A({\bf p})=\frac{{\vec \sigma}\cdot{\bf p}+i\epsilon_5}{\Omega_\tp+m}\tilde V_B({\bf p}),
\end{eqnarray}
which become, for helicity spinors, 
\begin{eqnarray}\label{cond_no_spinor_tilde}
\tilde U_B({\bf p})&=&\frac{2\lambda \tp+i\epsilon_5}{\Omega_\tp+m}\tilde U_A({\bf p}),\cr
\tilde V_A({\bf p})&=&\frac{-2\lambda\tp+i\epsilon_5}{\Omega_\tp+m}\tilde V_B({\bf p}).
\end{eqnarray}

We can impose now the boundary conditions at $t=0$,
\begin{equation}\label{NJL_no}
\Psi_\pm({\bf x},0)=\psi_\pm({\bf x},0),
\end{equation}
where $\psi_\pm({\bf x},0)$ are given by \eqref{Majorana_mode_exp}. By equating the solutions according to \eqref{NJL_no},  we expect to obtain the Bogoliubov transformations \eqref{Ansatz_1} for $\epsilon_1=0$, i.e.
\begin{eqnarray}\label{BT_no}
A_{\lambda}({\bf p})&=&\alpha_\tp a_{M\lambda}({\bf p})+\sgn\lambda\,\beta_\tp\,a^\dagger_{M\lambda}(-{\bf p}),\cr
B_{\lambda}({\bf p})&=&\alpha_\tp b_{M\lambda}({\bf p})+\sgn\lambda\,\beta_\tp\,b^\dagger_{M\lambda}(-{\bf p}),
\end{eqnarray}
with 
\begin{eqnarray}\label{BT_coeff_no}
\alpha_\tp=\sqrt{\frac{\Omega_\tp+\omega_\tp}{2\Omega_\tp}},\ \ \ \ \ \ 
\beta_\tp=-\sqrt{\frac{\Omega_\tp-\omega_\tp}{2\Omega_\tp}}.
\end{eqnarray}

In contrast to the preceding subsection, we shall adopt this time a bottom-up approach, proving the consistency of the boundary condition \eqref{NJL_no} with the Bogoliubov transformations \eqref{BT_no}. To this end, we start from $\psi_\pm({\bf x},0)$ written in terms of $a_{M\lambda}, a^\dagger_{M\lambda}$ and $b_{M\lambda}, b^\dagger_{M\lambda}$, as in \eqref{Majorana_mode_exp}. We consider the Bogoliubov transformations \eqref{BT_no} known, and apply their inverses
\begin{eqnarray}\label{BT_no_inv}
a_{M\lambda}({\bf p})&=&\alpha_\tp A_{\lambda}({\bf p})-\sgn\lambda\,\beta_\tp\,A^\dagger_{\lambda}(-{\bf p}),\cr
b_{M\lambda}({\bf p})&=&\alpha_\tp B_{\lambda}({\bf p})-\sgn\lambda\,\beta_\tp\,B^\dagger_{\lambda}(-{\bf p}).
\end{eqnarray}
The resulting solutions {\it have to} be $\Psi_\pm({\bf x},0)$, i.e. solutions of the equations of motion \eqref{eom_no_NJL}, if the boundary conditions \eqref{NJL_no} are compatible with the Bogoliubov transformations \eqref{BT_no}. We have only to confirm whether that is indeed the case, by verifying if the conditions \eqref{cond_no_spinor} and \eqref{cond_no_spinor_tilde} are fulfilled.

Let us proceed with $\psi_+({\bf x},0)$:
\begin{eqnarray}
\psi_+({\bf x},0)&=&\int\frac{d^3p}{(2\pi)^{3/2}\sqrt{2\omega_\tp}}e^{i{\bf p\cdot x }}\sum_\lambda\left(a_{M\lambda}({\bf p})u_\lambda({\bf p})+\sgn\,\lambda\ a^\dagger_{M\lambda}(-{\bf p})v_\lambda(-{\bf p})\right),\cr
&=&\int\frac{d^3p}{(2\pi)^{3/2}\sqrt{2\omega_\tp}}e^{i{\bf p\cdot x }}\sum_\lambda\Big[\left(\alpha_\tp A_{\lambda}({\bf p})-\sgn\lambda\,\beta_\tp\,A^\dagger_{\lambda}(-{\bf p})\right)u_\lambda({\bf p})\cr
&+&
\sgn\,\lambda\ \left(\alpha_\tp A^\dagger_{\lambda}(-{\bf p})+\sgn\lambda\,\beta_\tp\,A_{\lambda}({\bf p})\right)v_\lambda(-{\bf p})\Big],\cr
&=&\int\frac{d^3p}{(2\pi)^{3/2}\sqrt{2\omega_\tp}}e^{i{\bf p\cdot x }}\sum_\lambda\Big[\left(\alpha_ \tp u_\lambda({\bf p})+\beta_ \tp v_\lambda(-{\bf p})\right) A_{\lambda}({\bf p})\cr
&+&
\sgn\,\lambda\ \left(\alpha_\tp v_\lambda(-{\bf p})-\beta_\tp\,u_\lambda({\bf p})\right)A^\dagger_{\lambda}(-{\bf p})\Big],\cr
&=&\int\frac{d^3p}{(2\pi)^{3/2}\sqrt{2\Omega_\tp}}e^{i{\bf p\cdot x }}\sum_\lambda\left(A_{\lambda}({\bf p})U_\lambda({\bf p})+\sgn\,\lambda\ A^\dagger_{\lambda}(-{\bf p})V_\lambda(-{\bf p})\right),\cr
&=&\Psi_+({\bf x},0).
\end{eqnarray}
Thus, we identify
\begin{eqnarray}\label{UV_check}
U_\lambda({\bf p})=\sqrt{\frac{\Omega_\tp}{\omega_\tp}}\ \big(\alpha_ \tp u_\lambda({\bf p})+\beta_ \tp v_\lambda(-{\bf p})\big),\cr
V_\lambda({\bf p})=\sqrt{\frac{\Omega_\tp}{\omega_\tp}}\ \big(\alpha_\tp v_\lambda({\bf p})-\beta_\tp\,u_\lambda(-{\bf p})\big).
\end{eqnarray}
It remains now to check whether $U_\lambda({\bf p})$ and $V_\lambda({\bf p})$ defined by \eqref{UV_check} do satisfy the conditions \eqref{cond_no_spinor}. Let us verify this for $U_\uparrow({\bf p})$, using \eqref{helicity_spinors}, \eqref{helicity_spinors_minus} and \eqref{BT_coeff_no}:
\begin{eqnarray}\label{U_up}
U_\uparrow ({\bf p})&=&\sqrt{\frac{\Omega_\tp}{\omega_\tp}}\sqrt{\omega_\tp+m}\left[\sqrt{\frac{\Omega_\tp+\omega_\tp}{2\Omega_\tp}}\left(\begin{array}{c}
            \chi_\uparrow\\
            \frac{\text{p}}{\omega_\tp+m}\chi_\uparrow
            \end{array}\right)\ 
-i\sqrt{\frac{\Omega_\tp-\omega_\tp}{2\Omega_\tp}}\left(\begin{array}{c}
           -\frac{\text{p}}{\omega_\tp+m} \chi_\uparrow\\
            \chi_\uparrow
            \end{array}\right)\right]\cr
&=&\sqrt{\frac{\Omega_\tp+m}{2\Omega_\tp}}\left(\begin{array}{c}
            \left(\sqrt{\Omega_\tp+\omega_\tp}+i\frac{\tp}{\omega_\tp+m}\sqrt{\Omega_\tp-\omega_\tp}\right)\chi_\uparrow\\
            \left(\sqrt{\Omega_\tp+\omega_\tp}\frac{\tp}{\omega_\tp+m}-i\sqrt{\Omega_\tp-\omega_\tp}\right)\chi_\uparrow
            \end{array}\right).
\end{eqnarray}
This spinor is a solution of \eqref{Dirac_no_p}, provided that it satisfies \eqref{cond_no_spinor} for $\lambda=1/2$, i.e.
\begin{eqnarray}
\frac{\sqrt{\Omega_\tp+\omega_\tp}\frac{\tp}{\omega_\tp+m}-i\sqrt{\Omega_\tp-\omega_\tp}}{\sqrt{\Omega_\tp+\omega_\tp}+i\frac{\tp}{\omega_\tp+m}\sqrt{\Omega_\tp-\omega_\tp}}=\frac{\tp-i\epsilon_5}{\Omega_\tp+m}.
\end{eqnarray}
This equality is straightforwardly confirmed. All the other spinors defined by  \eqref{UV_check} are similarly proven to satisfy \eqref{cond_no_spinor}\footnote{The expression \eqref{U_up} justifies why we preferred to adopt the bottom-up approach: the form of the solution for which the Bogoliubov transformations \eqref{BT_no} are obtained would be very difficult to guess based only on the conditions \eqref{cond_no_spinor}.}.

By similar considerations starting from $\psi_-({\bf x},0)$, one proves that
\begin{eqnarray}\label{UVtilde_check}
\tilde U_\lambda({\bf p})=\sqrt{\frac{\Omega_\tp}{\omega_\tp}}\ \big(\alpha_ \tp u_\lambda({\bf p})-\beta_ \tp v_\lambda(-{\bf p})\big),\cr
\tilde V_\lambda({\bf p})=\sqrt{\frac{\Omega_\tp}{\omega_\tp}}\ \big(\alpha_\tp v_\lambda({\bf p})+\beta_\tp\,u_\lambda(-{\bf p})\big)
\end{eqnarray}
satisfy the corresponding equations of motion \eqref{cond_no_spinor_tilde}.

%%%%%%%%%%%%%%%%%%%%%%%%%%%%%%%%%%%%%5

\end{document}